\documentclass[]{fairmeta}

\title{ScaleAcross Explorer: Exploring Communication Optimization for Scale-Across AI Model Training}

\author[1,2*]{Minghao Li}
\author[1,2*]{Alicia Golden}
\author[2]{Samuel Hsia}
\author[2]{Michael Kuchnik}
\author[2]{Adi Gangidi}
\author[2]{Xu Zhang}
\author[2]{Ashmitha Jeevaraj Shetty}
\author[2]{Zachary DeVito}
\author[2]{Weiwei Chu}
\author[2]{Dong He}
\author[2]{Haoci Zhang}
\author[2]{Yuchen Hao}
\author[2]{Ruoming Pang}
\author[2]{James Hongyi Zeng}
\author[2]{Ying Zhang}
\author[1,2*]{Minlan Yu}
\author[2]{Carole-Jean Wu}

\affiliation[1]{Harvard University}
\affiliation[2]{Meta Platforms, Inc.}
\contribution[*]{Work done at Meta}

\usepackage{xspace}
\usepackage{amsmath}
\usepackage{array}
\usepackage{caption}
\usepackage{graphicx}
\usepackage{ragged2e}
\usepackage{subcaption}
\usepackage{algorithm}
\usepackage{algpseudocode}
\usepackage{xurl}

\definecolor{main}{HTML}{5989cf}    
\definecolor{sub}{HTML}{cde4ff}     

\tcbset{
    sharp corners,
    colback = white,
    before skip = 0.2cm,    
    after skip = 0.5cm      
}                           

\newtcolorbox{boxA}{
    fontupper = \bf,
    boxrule = 1.5pt,
    colframe = black 
}

\newtcolorbox{boxB}{
    boxrule = 1pt,
    colframe = main,
    rounded corners,
    arc = 5pt   
}

\algblock{Input}{EndInput}
\algnotext{EndInput}
\algblock{Output}{EndOutput}
\algnotext{EndOutput}
\newcommand{\Desc}[2]{\State \makebox[2em][l]{#1}#2}
\algnewcommand\algorithmicforeach{\textbf{for each}}
\algdef{S}[FOR]{ForEach}[1]{\algorithmicforeach\ #1\ \algorithmicdo}
\algrenewcommand{\algorithmiccomment}[1]{/* #1 */}

\newif\ifpptext
\pptextfalse
\ifpptext

\else

\fi

\newif\ifcomm
\commfalse
\ifcomm
\newcommand{\minghao}[1]{\textbf{\color{red}Minghao: #1}}
\newcommand{\minlan}[1]{\textbf{\color{blue}Minlan: #1}}
\newcommand{\alicia}[1]{\textbf{\color{teal}Alicia: #1}}
\else
\newcommand{\minghao}[1]{}
\newcommand{\minlan}[1]{}
\newcommand{\alicia}[1]{}
\fi

\newif\ifmain
\maintrue
\ifmain
\newcommand{\discuss}[1]{}
\newcommand{\todo}[1]{}
\else
\newcommand{\discuss}[1]{#1}
\newcommand{\todo}[1]{#1}
\fi

\newcommand{\sysname}{\textsc{ScaleAcross Explorer}\xspace}

\newcommand{\parab}[1]{\noindent\textbf{#1}}

\newif\ifupdate
\updatefalse
\ifupdate
\newcommand{\newchange}[1]{{\color{cyan}#1}}
\else
\newcommand{\newchange}[1]{#1}
\fi

\abstract{The rapid scaling of large language model training requires distributing GPU resources across multiple data center buildings and regions. We refer to such paradigm as ``scale-across'' training. As infrastructure expands, the system design space becomes increasingly intricate, encompassing new model architectures, hardware heterogeneity, and evolving communication patterns. Drawing from Meta’s production experience, we highlight the complexities of deploying training jobs across a few data centers housing hundreds of thousands of GPUs. To accelerate exploration of the large design space and to enable efficient training for frontier model development, we conduct in-depth characterization of three key design dimensions: parallelism placement, parallelism scheduling, and network layer technologies. We then propose ScaleAcross Explorer --- an optimizer that considers the interplay of design dimensions and holistically optimizes scale-across training. Testbed experiments and simulations demonstrate up to 64.62\% training speedups over production configuration and up to 37.59\% training speedups over state-of-the-art baseline across a wide range of design points.}

\date{\today}
\correspondence{Minghao Li at \email{minghaoli@g.harvard.edu}, Minlan Yu at \email{minlanyu@g.harvard.edu}, Carole-Jean Wu at \email{carolejeanwu@meta.com}.}

\begin{document}

\maketitle

\section{Introduction}
\label{intro}

Model scaling has emerged as a primary driver to advance machine intelligence. The size of pretraining jobs in terms of GPUs required is rapidly growing.
As the scale of AI training infrastructure grows,
it becomes increasingly impractical to concentrate all compute resources in one data center building or even a single geographic region~\citep{ml_dc_network_research, semianalysis_grid_power_blackout, google_resiliency, microsoft_geo, sailor}.


In this paper, we share the key learning and challenges from our production experience at Meta, where we successfully deploy training jobs across zones and buildings housing hundreds of thousands of GPUs~\citep{meta2025infrastructure}. Such multi-building training is enabled by advanced parallelism placement, scheduling, and networking protocols, which we delve into with detail in Section~\ref{background:experience_100K}.

As we continue to grow the training infrastructure across data center buildings with different distances and even across regions, the design space across the entire system stack becomes highly complex, including new model architectures, hardware heterogeneity (from the perspective of both accelerators and networks), and new communication challenges for distributed training at-scale. For example, emerging mixture-of-experts (MoE) models  exposes new parallelism opportunities to overlap communication latency with useful computations. 
Scale-across model training introduces new communication patterns influenced by parallelism strategies and infrastructure design choices. Data center buildings may be co-located within a region ($\sim$1:3-1:8 oversubscription and 10's-100 km distance) or distributed across regions (beyond 1:8 oversubscription and 100's km distance). As we show in this paper, training iteration time vary with cross-building bandwidth oversubscription and latency.


\begin{table}[t]
\caption{\small Characteristics of Scale-out (cross zones) vs Scale-across (cross buildings) networks.}
\label{table:cross_layers}
\centering
\begin{tabular}{|c|c|c|c|}
\hline
               & {\small Bandwidth} &  {\small Distance } & {\small Latency} \\ 
               & {\small Oversubscription} & {\small (km)} & {\small ($\mu$s)} \\
               
               \hline
{\small Cross Zones}     & {\small 1:2-1:4} & {\small $\sim$5}               & {\small 20-30}  \\ \hline
{\small Cross Buildings} & {\small 1:3 and beyond}  & {\small O(10)-O(100)} & {\small $\geq$ 50}  \\ \hline
\end{tabular}
\end{table}

\begin{figure*}[t]
    \centering    \includegraphics[width=0.9\linewidth,trim={2cm, 5cm, 1.5cm, 5.1cm}, clip]{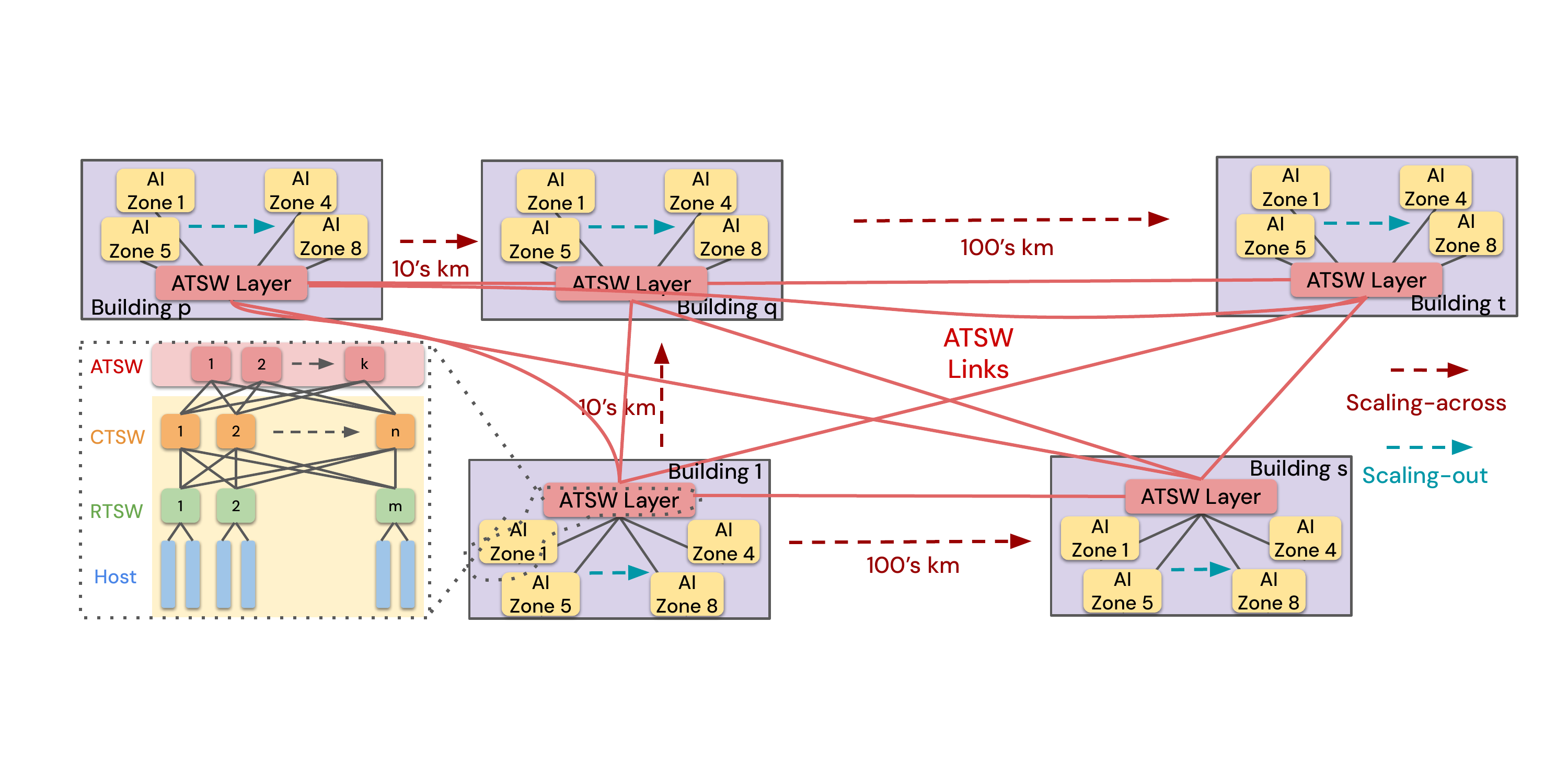}
    \caption{\small Scale-across model training infrastructure~\citep{ncclx_arxiv,atscale2025metadc}. RTSW: Rack Training Switch; CTSW: Cluster Training Switch; ATSW: Aggregation Training Switch.
    }
    \label{fig:scale_across_infra}
\end{figure*}

\begin{figure*}[t]
    \centering
    \includegraphics[width=0.97\textwidth,trim={5cm, 5.9cm, 4cm, 4.5cm}, clip]{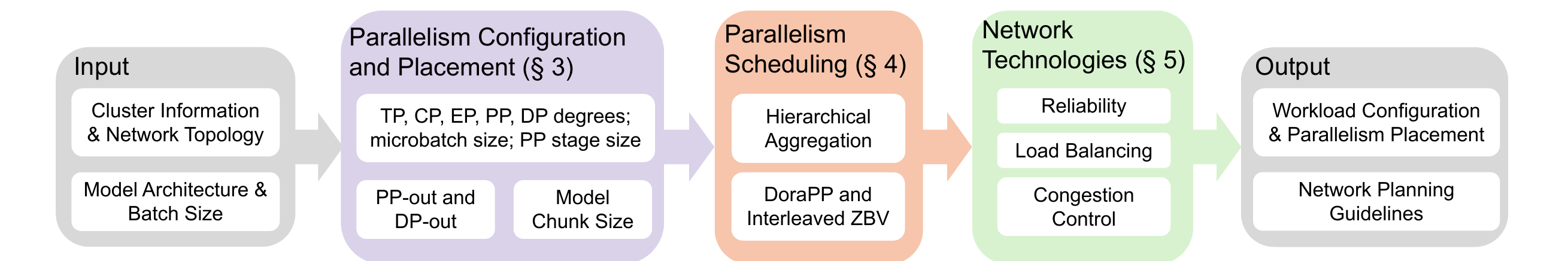}
    \caption{\small Overview for \sysname and the key design dimensions for scale-across model training.}
    \vspace{-0.25cm}
    \label{fig:optimizer_overview}
\end{figure*}

Figure~\ref{fig:scale_across_infra} provides an overview for Meta's scale-across model training infrastructure. We define scale-across (red dashed arrows in Figure~\ref{fig:scale_across_infra}) as model training expanded across multiple data center buildings, interconnected via cross-building networks. 
This contrasts with ``scale-out'' (teal dashed arrows), which increases capacity within a zone or connects additional zones through a homogeneous network.

To accelerate exploration of the large model design space and to enable training productivity for frontier model development, we deployed a scale-across testbed with configurable latency and bandwidth across clusters to comprehensively characterize the extensive design space of scale-across training.
While the design space characterization provides an in-depth understanding of key design dimensions of scale-across training, it does not consider the interplay of model architectures, parallelism placement and schedules, and network layer technologies. To optimize scale-across model training holistically, we propose \sysname---a new optimizer for scale-across training configuration. \sysname takes model architecture specification, batch sizes, network topology configurations, and hardware specifications as input. It then searches for the configuration across the stack of parallelism placement, parallelism schedules, and network protocols that minimizes iteration time. Figure~\ref{fig:optimizer_overview} provides an overview for the key design dimensions and the optimizer workflow.


We characterize the following key design dimensions in \sysname: 

\parab{Parallelism Placement.} In scale-across training, the choice of whether to place data parallelism or pipeline parallelsm as the outermost layer (DP-out vs PP-out) depends on model density and bandwidth oversubscription. For example, DP-out uses the cross-building links less frequently and thus performs better for dense model. For MoE models, as the number of experts increases, DP data volume grows signficantly, making PP-out a better choice.


\parab{Parallelism Scheduling.} In production, there are many variations on DP and PP scheduling strategies. For example,  Hybrid Sharded Data Parallel (HSDP) is an alternative to Fully Sharded Data Parallelism (FSDP)~\cite{fsdp} to mitigate scale-across network bottlenecks. We have also extended flexible PP~\cite{llama3} to DoraPP and Interleaved Zero Bubble V-schedule (ZBV) with different tradeoffs between computation efficiency and cross-building traffic. The choice of these solutions depend on cross-building network settings.


\parab{Network Layer.} The measurement on our scale-across testbed shows that collective communication times depend on message sizes and cross-building distances. The performance of DP and PP traffic on across-building links also depends on latency, loss rates, and transport configurations (e.g., ECMP vs packet spraying). 


\sysname automatically navigates the large design space and identifies cross-training configurations that achieves 16.38\% to 37.59\% speedup over the state-of-the-art baseline for various model and network settings. 

\section{Deployment Experience for Scaling-Across Training}
\label{background}


The ever-increasing scale of large language model training demands innovations in infrastructure design. 
Llama 3 was trained with 16K GPUs located in one data center building~\citep{llama3, meta_dc_networks}. With Llama 4, the training scale reached 100K+ GPUs across data center buildings~\citep{meta_dc_networks}. In this section, we share our experiences running cross-building training jobs. We then discuss the challenges of the complex design space as we further scale training jobs, and show our new scale-across testbed for characterizing the design choices.


\subsection{Cross-Building Training Configurations} 
\label{background:experience_100K}

\parab{Infrastructure.} Our 100K-GPU deployment spans across multiple buildings (Figure~\ref{fig:scale_across_infra}). The cross-building network interconnects the aggregation-layer switches (ATSW) of each building, with bandwidth oversubscription ratio in the range of $\sim$1:3\citep{meta_dc_networks} to 1:8 and latencies rising to $4\times$ longer latency than intra-zone network~\citep{meta_100k_optimization}. 
The bandwidth and latency of cross-building links in this infrastructure is at the lower end of the numbers in Table~\ref{table:cross_layers}.
Each building contains multiple AI zones, supporting tens of thousands of GPUs.
These zones are interconnected via a Clos architecture with three layers of switches (ATSW, cluster-layer switches (CTSW), and rack-layer switches (RTSW))~\citep{ncclx_arxiv}.






\parab{Parallelisms placement.}
Our cross-building training jobs encompass five dimensions of parallelisms: Tensor Parallelism (TP), Context Parallelism (CP), Expert Parallelism (EP), Pipeline Parallelism (PP), and Data Parallelism (DP). We place DP on the outermost cross-building layer and place PP on the cross-zone layer.  
Due to their high communication overhead, TP, CP, and EP require high bandwidth and low latency network and are confined within AI Zone.

DP partitions the training dataset across GPUs. We employ Fully Sharded Data Parallelism (FSDP)~\citep{fsdp}, which shards both model parameters and gradients among workers. By applying DP at the outermost layer, gradient ReduceScatter and parameter AllGather operations occur over the cross-building network. Gradient synchronization for each layer can begin immediately after its backward pass and overlap with the backward computation of earlier layers; the same applies to parameter prefetching~\citep{llama3}. Therefore, we identify DP as tolerant to low bandwidth and long latency of the cross-building network.

We place PP as the second-to-outermost parallelism on the cross-zone network. PP divides model layers into sequential stages and introduces point-to-point communication of activation and activation gradient between adjacent stages per model chunk per microbatch. PP demands higher bandwidth and is less tolerant of long-distance communication than DP. Any PP communication that is not fully overlapped with computation becomes part of the critical path.

\begin{figure}[t]
    \centering
    \includegraphics[width=0.95\hsize,trim={5.6cm, 2.7cm, 4.1cm, 8.9cm},clip]{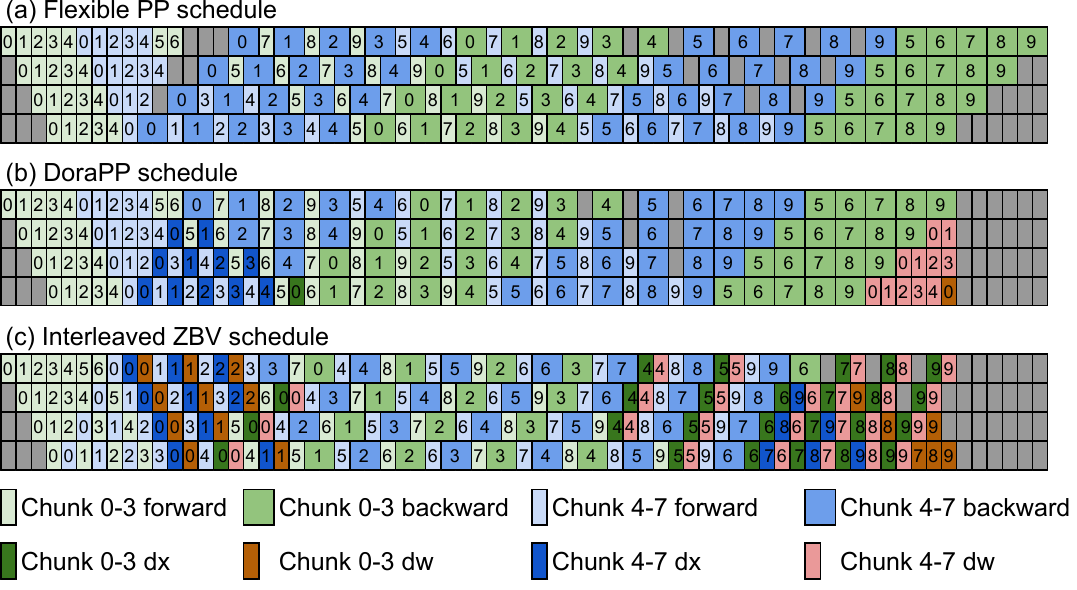}
    \caption{\small \small A comparison of different pipeline schedules using an example setup with 4 PP ranks, 2 model chunks per PP rank (8 chunks in total), and 8 microbatches: (a) Llama 3 Flexible PP~\cite{scaling_llama3}, (b) DoraPP, (c) Interleaved ZBV~\cite{zerobubble}. Within each pipeline schedule, each row represents a PP rank, each column represents a time unit and dashed lines show propagation of the first microbatch.}
    \label{fig:pp_schedule}
    \vspace{-0.25cm}
\end{figure}

\parab{Parallelism scheduling.}
We inherit parallelism schedule choices for intra-building training. For DP, we use FSDP. For PP, we extend the flexible PP schedule used in Llama 3~\citep{scaling_llama3} to Doraemon PP (DoraPP) to further reduce bubbles (Figure~\ref{fig:pp_schedule}). Inspired by the zero bubble schedule~\citep{zerobubble},  DoraPP splits backward computation into input (dx) and weight (dw) gradients to maximize overlap between computation and communication. Like flexible PP, DoraPP uses a ``wrap-around'' communication pattern, connecting the last and first pipeline stages for adjacent model chunks. DoraPP executes the same layer for multiple microbatches on every rank, allowing prefetching weights for the next layer and thereby enabling integration with ZeRO-2 and ZeRO-3~\citep{zero_infinity}, which reduces gradient and weight memory usage.




\parab{Network layer.} As discussed in our previous work on the operation of RDMA for distributed training~\citep{meta_roce_at_scale}, we disable congestion control because our collective-library level congestion management, combined with deep-buffer spine switches (CTSW and ATSW), already ensures stable training performance without persistent congestion.
Overall, 11.3\% of total traffic is cross-building, and 0.05\% and less than 0.01\% of packets are ECN-marked for intra- and cross-building communication, respectively. The 99th percentile latencies are 40~$\mu$s for intra-building and 85~$\mu$s for cross-building communication.


With these configurations, we have been successfully running cross-building training jobs at 100K GPU scale with high GPU utilization and network stability since Llama 4 pretraining~\citep{meta_dc_networks, meta_100k_optimization}.

\subsection{Complex Scale-Across Design Space}
\label{background:scale-across-prod-choices}

Building on our experiences of running cross-building training, we aim to further scale training jobs to support next-generation large models requiring millions of GPUs. However, as gigawatt-scale energy demands and grid constraints prevent single locations from supporting hundreds of thousands of synchronous GPUs~\citep{ml_dc_network_research,semianalysis_grid_power_blackout,ai_power_stabilization}—we are compelled to expand to multiple buildings across distributed regions. In this expanded topology, inter-building distances are dictated by power, grid, and cooling constraints, resulting in significant variability (Table~\ref{table:cross_layers}). For example, regions may be separated by more than 200 kilometers, introducing bandwidth oversubscription of 1:8 or higher and latencies ranging from 100~$\mu$s to several milliseconds~\citep{ml_dc_network_research,semianalysis_multidc,tomshardware2024chinaai}. 
This variability necessitates us revisiting the configurations for our scale-across training jobs from Section 2.1. We only discuss the design options that are practical to implement at today's devices and make significant impact on the iteration time from our production experiences.


\parab{Parallelisms placement.} Although our cross-building training uses DP-out, previous Microsoft and academic work~\citep{microsoft_geo,euromlsys_cross_region,sailor,crosspipe} suggested PP-out for geo-distributed training. As MoE models introduce new parallelism options and communication patterns, we need to revisit DP-out and PP-out for more model and network configurations.

\parab{Parallelisms scheduling.} As cross-building links have longer delay and lower bandwidth, we need to revisit those parallelism solutions that can reduce traffic on these links. In practice, we only consider lossless solutions to avoid affecting ML team optimization and debugging.
For DP scheduling, hierarchical communication (e.g., Hybrid Sharding Data Parallel, HSDP)\citep{fsdp} combines inner-group parameter sharding with model replication across groups. We can place replica groups across building, so that bandwidth-intensive parameter AllGather stays intra-building and we only perform gradient synchronization across buildings.
 

For PP scheduling, although we chose DoraPP for intra-building settings, we revisit other alternatives with different computation-communication tradeoffs because communication overhead becomes more important for cross-building settings. For example, alongside DoraPP, we invented Interleaved Zero Bubble V-schedule (ZBV in Figure~\ref{fig:pp_schedule}) separates the calculation of input gradients from weight gradients. The schedule prioritizes the input gradients to immediately unblock upstream devices, while deferring the weight gradient computations. These deferred tasks are then executed during the gaps that would normally result in pipeline bubbles, reducing activation memory usage. Compared to DoraPP, interleaved ZBV incurs more computation overhead due to its more aggressive backward splitting. But its V-schedule enforces a non-cyclic dependency chain, eliminating the wrap-around communications between the last and first stages, and thus reducing cross-building communications.

\parab{Network layer.} For our 100K-GPU cross-building training, we inherited intra-building network protocols described in our prior paper~\citep{meta_roce_at_scale}. However, as the cross-building distance further grows, it becomes necessary to systematically characterize latency, bandwidth, and packet loss behavior, since these factors directly influence future protocol design choices. For example, we need to revisit the utility of selective redundancy for proactive loss mitigation, the throughput comparison between ECMP and packet spraying, and the effectiveness of ECN-based congestion control.

\subsection{Scale-Across Testbed}
\label{method_testbed_simulation}

To systematically characterize the impact of configurations for cross-building training, we build a scale-across testbed in production. Our testbed features multiple independent clusters, each equipped with multiple racks of H100 GPUs. These clusters are fully meshed via optical systems allowing flexible configurations of cross-building topology and bandwidth. We also introduced  
fiber spools of up to 500km (similar to the M2 Optics Fiber Lab~\citep{m2optics_fiberlab}) to support configurable distances between buildings.

\begin{table}[t]
\caption{\small Workload configurations for the testbed experiments and emulation (8K sequence length). Mbs: ``Microbatch size''; ``Gbs'': ``Global batch size''. }
\label{table:workload_configs}
\centering
\begin{tabular}{|c|c|c|c|c|}
\hline
 Model & Mbs, Gbs & TP-CP-EP-PP-DP & Chunk size & Experts \\ \hline
 40B-MoE & 1, 64 & 4-1-16-2-16 & 2 layer & 16-128\\ \hline
 17B-Dense & 4, 176 & 8-1-NA-2-4 & 1 layer & NA \\ \hline
\end{tabular}
\end{table}

We configure bandwidth oversubscription ratios of 1:1.33, 1:4, 1:8, and 1:16, encompassing all scenarios described in Table~\ref{table:cross_layers}. To achieve different oversubscription ratios on the production testbed, we incrementally drained uplinks on the CTSW. Out of the 48 links between the 2 clusters, we drained 32, 40, and 44 links to obtain 1:4, 1:8, and 1:16 oversubscription ratios respectively.
In production, adjusting port speed with software rate limiters such as ``ethtool'' is not feasible due to auto-negotiation of link speed between senders and receivers. \minlan{why do we have auto negotiation?}If the sender changes link speed, the receiver will detect the mismatch and disable the link. Unless otherwise specified, the default oversubscription ratio on the production testbed is 1:1.33, which is the minimum achievable given the testbed configuration.

We choose dense and MoE models based on our production experiences and list their configuration in Table~\ref{table:workload_configs}. 
Because the production testbed also supports other users, we prioritize experiments with MoE models on scale-across testbed given our limited time slots, but run the dense model experiments on another research cluster of 64 H100 GPUs (see Appendix~\ref{app:method_emulator} for detailed description).
To validate our findings at 100K-GPU scale, we employ an in-house packet-level network simulator~\citep{arcadia} which are validated and used in production. This simulator further enable characterizing the effects of link latency, packet loss rates, and network protocols. 
Simulations are configured using model architectures and workload settings mirroring the 100K GPUs training job we discussed in Section~\ref{background:experience_100K}. We simulate multiple data center buildings and cover 1:1, default, and 2$\times$ and 4$\times$ the default oversubscription ratios (``default ratio'' is based on that of an internal multi-building region), as well as link latencies of 10, 25, 50, 100, and 1000 $\mu$s, with 50 $\mu$s being the default. The default topology is designated as ``Topology A1'' in the later text. To demonstrate the generalizability of our findings, we further simulate multiple topology variants (configurations in Table~\ref{table:topology_configs}). Simulation setup details are in Appendix~\ref{app:method_simulator}.

In the following sections, we share our characterization of parallelism placement, parallelism scheduling, and network layer technologies with extensive experiments on our scale-across testbed and large-scale simulations.


\section{Parallelism Placement}
\label{sec:parallelism_schemes}
When configuring distributed training across long distances, the key initial decision is which parallelism strategy to place at the outermost layer, as this choice shapes all subsequent optimizations. We analyze the performance factors of PP-out versus DP-out. While prior geo-distributed studies often favor PP-out, our findings show that PP-out can be less effective for dense model training.

\subsection{Dense Models} 
\label{emulaion:dp_pp_optimal_batch_size}


Let $H$ be model dimension, $S$ be sequence length, $m$ be microbatch size, $F$ be Feed-Forward Networks (FFN) dimension, $E$ be number of experts, and $F_e$ be expert intermedaite dimension. Then for each iteration, the data transmitted in point-to-point (P2P) communication between adjacent PP stages is roughly $2 \times \frac{H}{\text{TP}} \times \frac{S}{\text{CP}}$ $\times m$, with a total of $\text{microbatches} \times \text{model chunks}$ such communications in each iteration. With DP-out, the per-layer data exchanged per iteration is approximately $\frac{1}{\text{TP}}\times$ $(4\times H^2$ $+ 3\times H \times F)$ for Llama-style dense models, and $\frac{1}{\text{TP}} \times$ $(4\times H^2$ $+3\times H$ $\times$ $F_e$ $\times \frac{E}{\text{EP}})$ for MoE models. As discussed in Section~\ref{background:experience_100K}, with PP-out, P2P communications between stages may cross building boundaries and become exposed. In contrast, DP-out introduces delay in parameter AllGather and gradient ReduceScatter per layer, but only the first layer’s communication is fully exposed.

\parab{Parallelism placement with varying bandwidth oversubscription.}
\label{subsubsection:dense_model_placement}
\begin{figure*}[t]
    \begin{minipage}[b]{0.46\textwidth}
      \centering 
      \includegraphics[width=\linewidth]{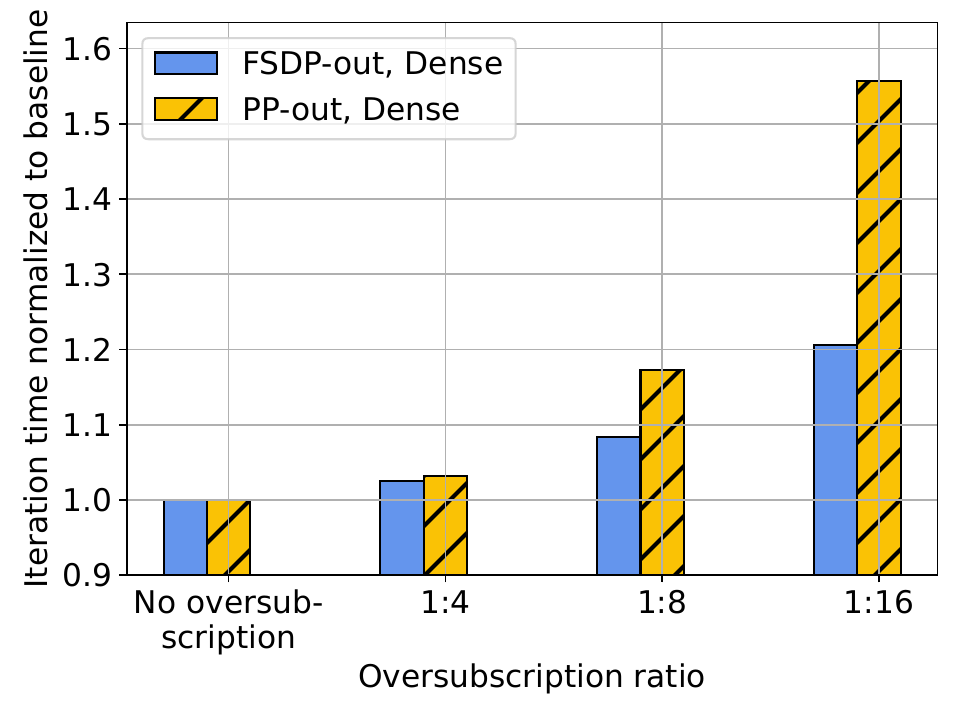}  
      \caption{[Testbed] 17B dense model iteration time, normalized to no oversubscription baseline.}
      \label{fig:fsdpout_ppout_gbs176}
    \end{minipage}
    \hfill
    \begin{minipage}[b]{0.46\textwidth}
      \centering 
      \includegraphics[width=\linewidth]{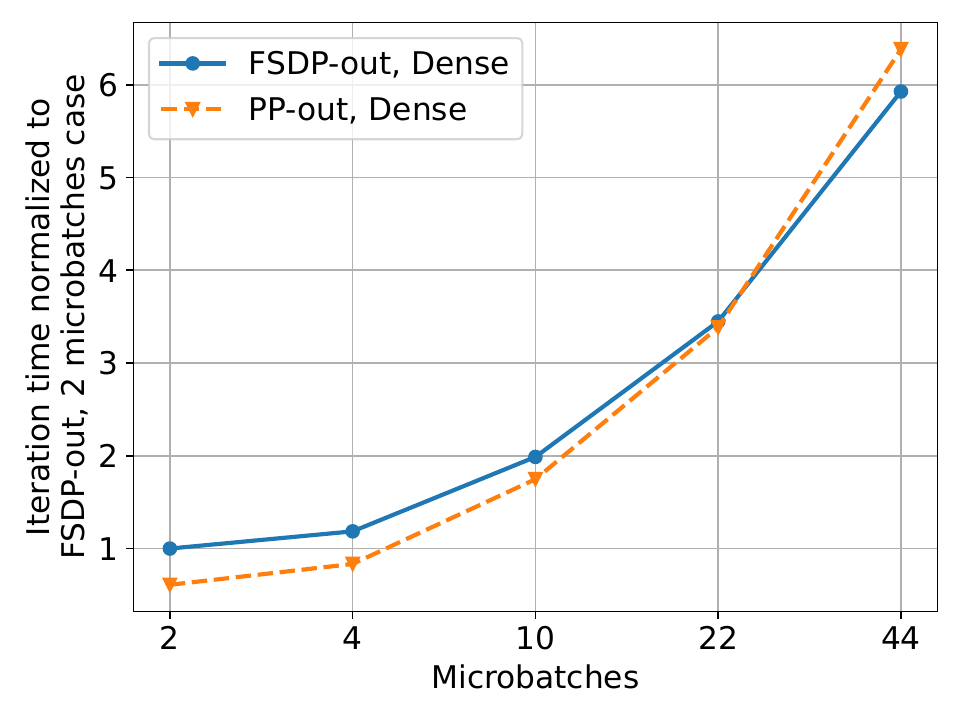} 
      \caption{[Testbed] Iteration time for PP-out and DP-out across varying number of microbatches. 
      }
      \label{fig:ppout_dpout_batchsize}
    \end{minipage}
    \vspace{-0.25cm}
\end{figure*}
As network oversubscription increases from cross-building to cross-region scenarios, how does the performance gap between PP-out and DP-out change? To answer this, we evaluate both the baseline case with no oversubscription (1:1) and emulated scenarios where inter-building GPU communication experiences oversubscription ratios of 1:4, 1:8, and 1:16. We set the global batch size as 176, which is training loss optimal following the LLM scaling law~\citep{batchsize_scaling}. 
As shown in Figure~\ref{fig:fsdpout_ppout_gbs176}, the performance gap between PP-out and DP-out is narrow when the oversubscription ratio is low. The critical threshold occurs at a 1:4 ratio. Beyond this point, DP-out is notably faster for the 17B dense model, with a 28.47\% speedup at a 1:16 oversubscription ratio. 

\begin{figure}[t]
\centering 
      \includegraphics[width=0.5\textwidth]{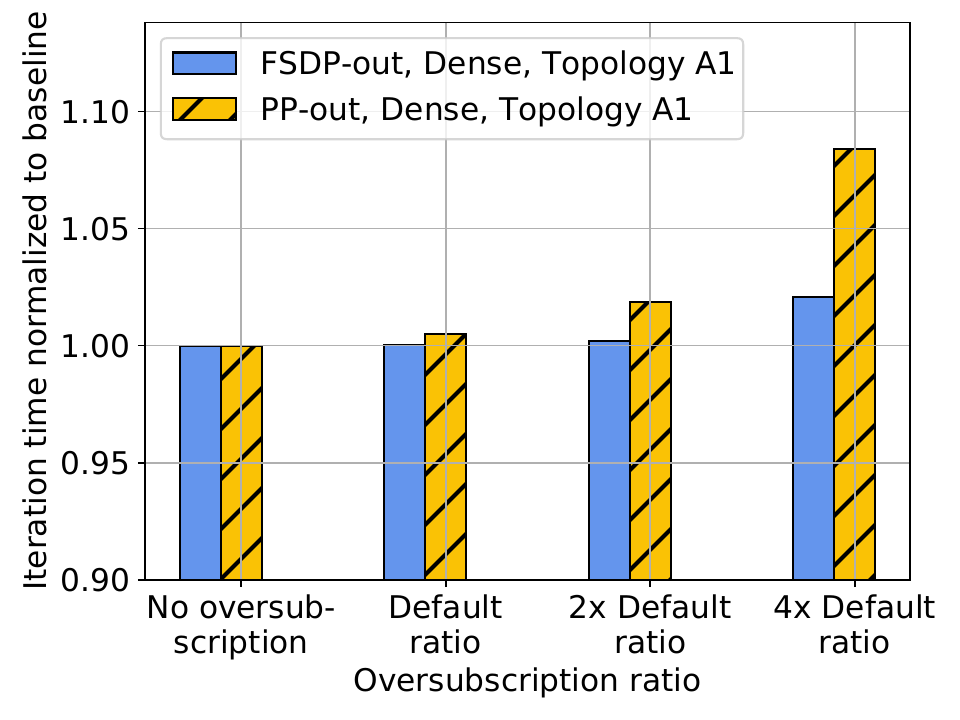} 
      \caption{\small [Simulation] Production trace based simulations.} 
      \vspace{-0.3cm}
      \label{fig:dpout_ppout_scale}
\end{figure}

\parab{Parallelism Placement with Varying Number of Microbatches.} In real-world deployments, batch sizes are set also considering the limitations of training tokens or compute resources and convergence speed. Here, we show that picking a smaller batch size, which leads to fewer microbatches in total, has a fundamental impact on PP-out performance. In this set of experiments, we gradually decrease the batch size from 176 to 8 and study the performance trend of DP-out and PP-out as we vary the batch size while keeping micro-batch size static as 4 (i.e., the number of microbatches equals $\frac{\text{batch size}}{4}$).

As shown in Figure~\ref{fig:ppout_dpout_batchsize}, with the 17B dense model, while PP-out is slower than DP-out under batch size 176, the performance gap closes when batch size is decreased to 88. As we further decrease the batch size, PP-out overtakes DP-out. The root cause behind the performance difference is the communication frequency. Pipeline parallelism involves pair-to-pair activation and gradient communication for each model chunk. When PP is the outermost parallelism, GPUs running adjacent stages might reside in different buildings, and therefore cross-building communication happens twice per model chunk per micro-batch (one for activation in the forward pass and another for gradient in the backward pass). On the other hand, when FSDP is the outermost parallelism, cross building communication only happens twice per iteration (one for parameter all-gathering and one for gradient reduce-scattering). In summary, a major performance determinant of PP-out is the number of micro-batches. Under a static parallelism configuration and an increasing number of micro-batches, the total volume of data communicated over cross-building links remains static for DP-out, but scales linearly for PP-out. When the number of micro-batches is large, PP-out transmits much more data over the cross-building links than DP-out, leading to worse performance at high oversubscription ratios.

\parab{PP-out vs. DP-out large scale simulations.}
\label{simulation:pp_vs_dp}
We conduct large-scale simulations using production workload traces, where 100K GPUs are placed into multiple buildings under a Clos network topology~\citep{ocp_summit_2025}. As in Figure~\ref{fig:dpout_ppout_scale}, we observe that PP-out is comparable or worse than DP-out (up to 6.60\% slower at the $4\times$ default oversubscription ratio). Two key factors contribute to this result. First, to prevent stage idling, it is necessary to ensure that the number of micro-batches per iteration at least equals the number of pipeline stages. As the production workload has many pipeline stages, the number of microbatches per iteration is large. Second, the model chunk size setting and pipeline schedule exacerbates the issue. The workload sets one single layer per model chunk and uses DoraPP, which assigns model chunks sequentially across pipeline stages in a round-robin fashion (e.g., with 6 model chunks and 2 pipelines stages, chunk 0, 2, and 4 are on stage 0, and chunk 1, 3, and 5 are on stage 1). This leads to hundreds of cross-building activation and gradient communication within one iteration. Consequently, PP-out exhibits a longer iteration time due to the high cross-building communication frequency.

\begin{figure*}[t]
    \begin{minipage}[b]{0.47\textwidth}
      \centering 
      \includegraphics[width=\linewidth]{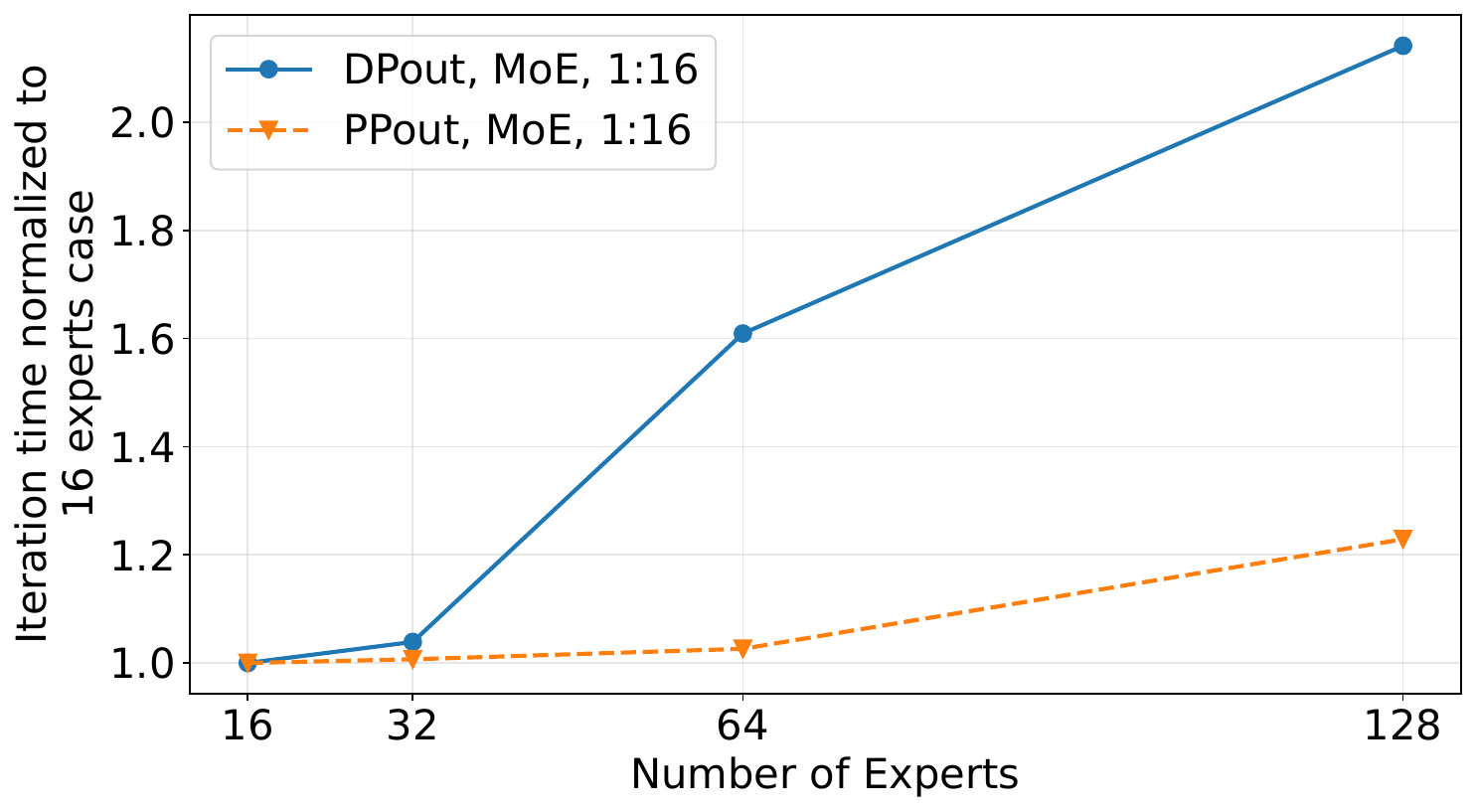} 
      \vspace{-0.25cm}
      \caption{\small [Testbed] Comparison of the iteration time with increasing number of experts for the 40B MoE model. DP-out iteration time increases faster than PP-out as the number of experts increases. 
      }
      \label{fig:moe_experts_sweep} 
    \end{minipage}
    \hfill
    \begin{minipage}[b]{0.45\textwidth}
      \centering 
      \includegraphics[width=\linewidth]{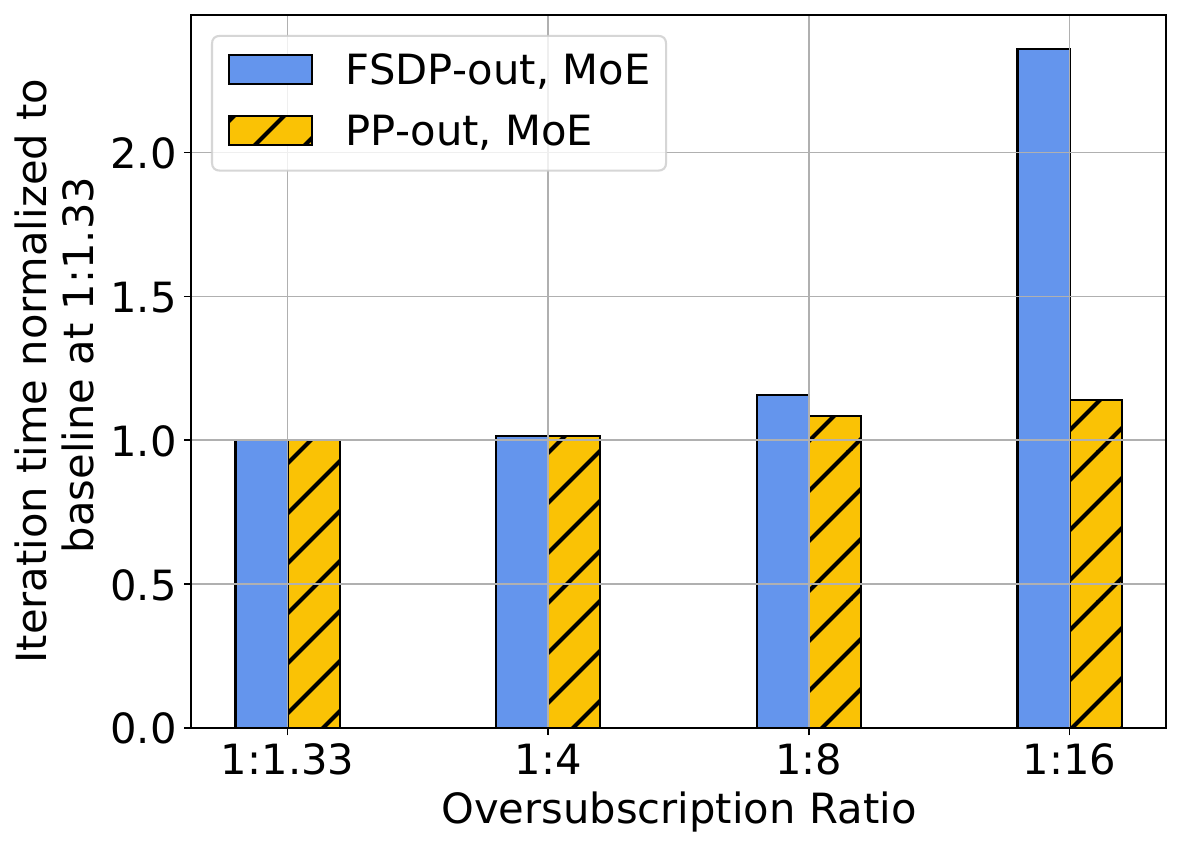}  
      \vspace{-0.25cm}
      \caption{\small [Testbed] 40B MoE model iteration time, normalized to baseline of 1:1.33 oversubscription ratio. PP-out yields faster iterations when the oversubscription ratio is larger than 1:4.}
      \label{fig:fsdpout_ppout_moe}
    \end{minipage}
\end{figure*}

\begin{boxB}
\textbf{Takeaways:} DP-out is better for training dense models with a large number of microbatches.
\end{boxB}

\subsection{MoE Models}
\label{subsec:moe_eval}
\parab{Parallelism placement with varying bandwidth oversubscription.}
The ongoing trend of model architecture design is leveraging Mixture-of-Experts (MoE) layers. To gather more realistic insights on scaling-across MoE model training, we conduct experiments on the production testbed with 128 H100 GPUs. Contrary to dense model results, Figure~\ref{fig:moe_experts_sweep} and Figure~\ref{fig:fsdpout_ppout_moe} show that PP-out scales better as the number of experts and oversubscription ratio increases. At a 1:16 oversubscription ratio with 128 experts, PP-out achieves a 49.4\% speedup over DP-out. PP-out scales better with MoE models because PP communication (activation and activation gradient) volume remains constant as the number of experts increases. In contrast, DP communication (model parameters and model gradients) volume grows linearly with the number of experts.

\begin{figure}[t]
    \centering
    \begin{subfigure}[h]{0.47\textwidth}
        \centering
        \includegraphics[width=\linewidth]{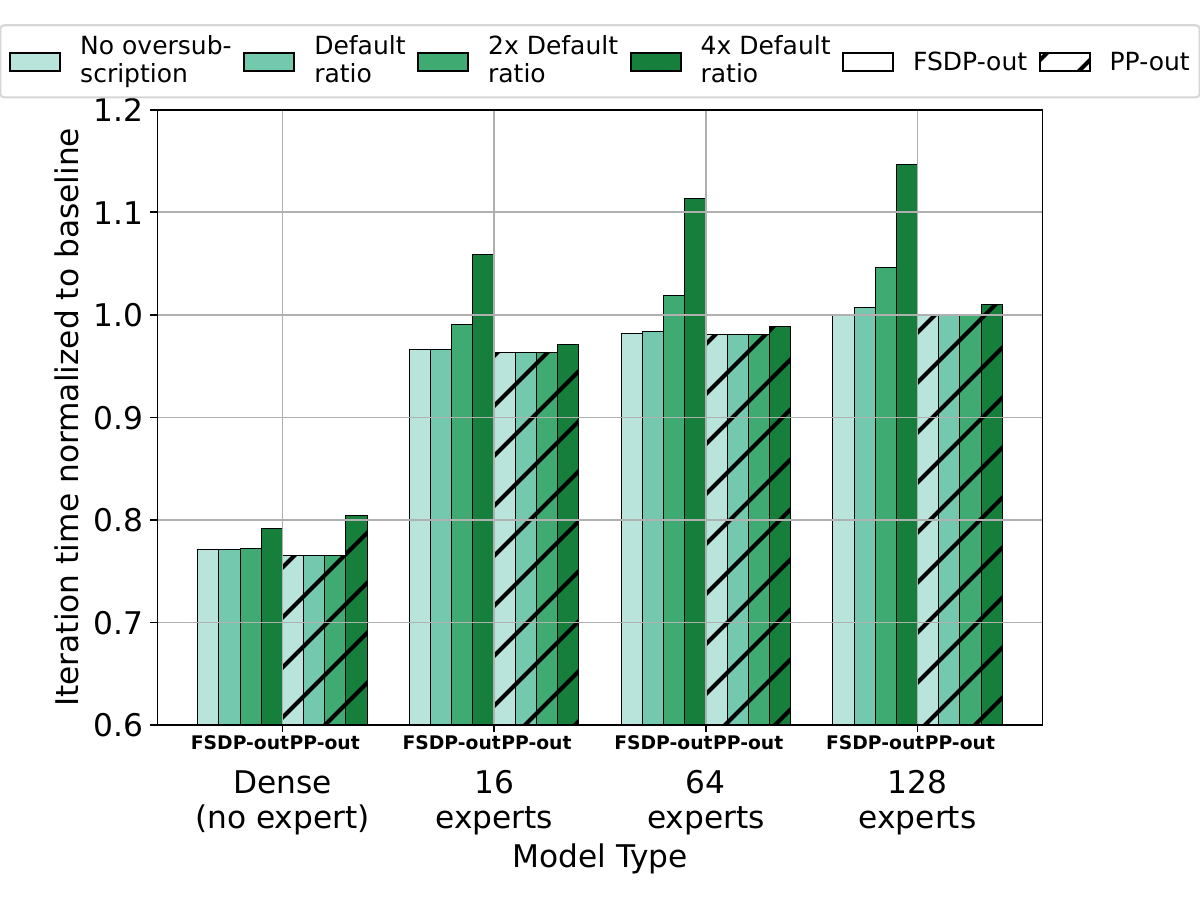}
        \vspace{-0.5cm}
        \caption{\small Impact of varying bandwidth oversubscription.}
        \label{fig:moe_bdw_pp16}
    \end{subfigure}
    \begin{subfigure}[h]{0.47\textwidth}
        \centering
        \includegraphics[width=\linewidth]{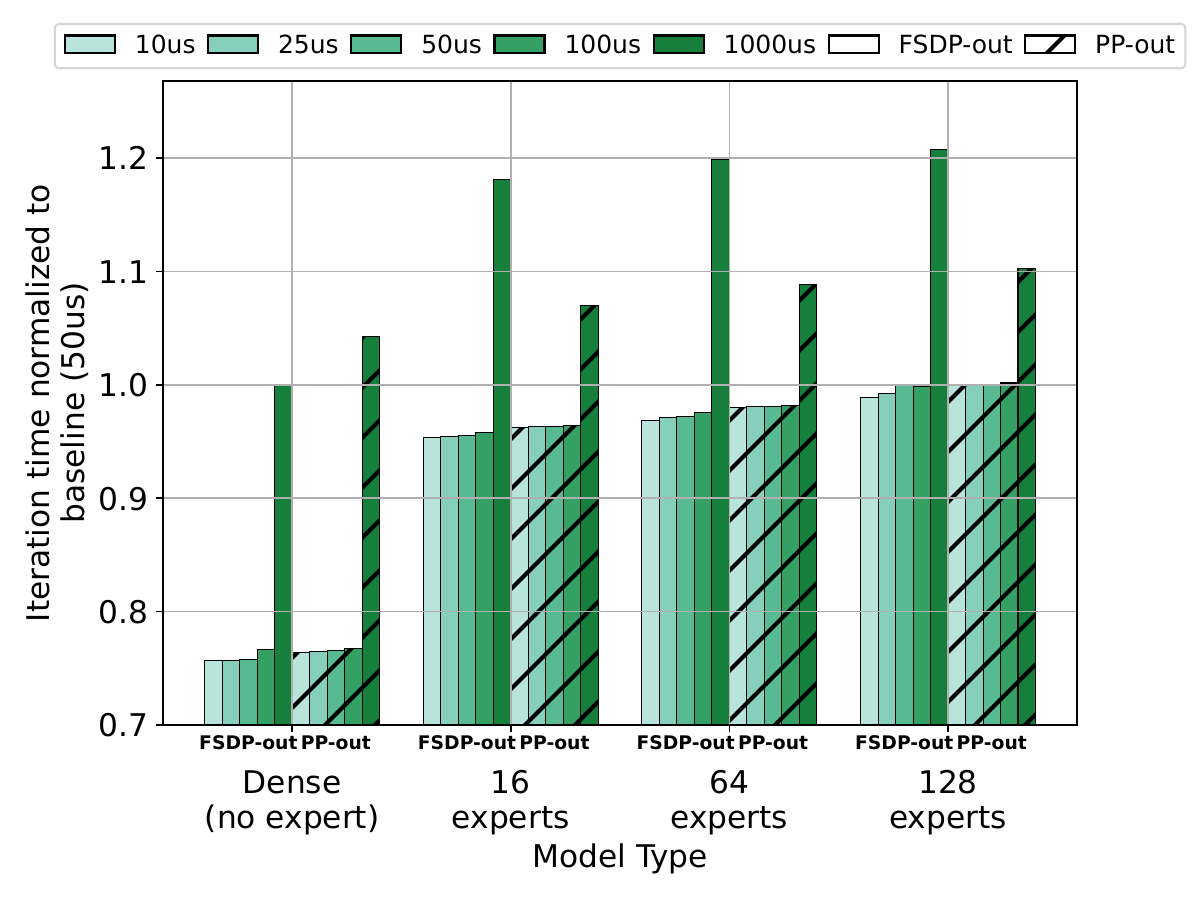}
      \vspace{-0.5cm}
        \caption{\small Impact of varying link latencies.}
        \label{fig:moe_latency_pp16}
    \end{subfigure}
    \vspace{-0.25cm}
    \caption{\small [Simulation] Training time comparison between MoE models with varying number of experts and dense model, under parallelism configuration TP4-PP16-DP128.
    }
    \label{fig:moe_vary_exps_pp16}
\end{figure}

\parab{Large scale simulations with varying experts.}
We conduct large-scale simulations for MoE model training by adapting the traces of an 8K GPUs production workload. We simulate two buildings, each hosting 4K GPUs. We configure TP degree as 4, PP degree as 16, and DP degree as 128 for the first set of experiments. As shown in Figure~\ref{fig:moe_latency_pp16}, increasing oversubscription ratios leads to substantial iteration time increase for MoE models using DP-out. However, high oversubscription ratio has limited impact on PP-out. Dense model iteration time increases at a high oversubscription ratio (4$\times$ default) when using either DP-out or PP-out. PP-out increases slightly more due to the shorter computation time of the dense model not hiding the longer cross-building P2P SendRecv time.

Longer latency increases the iteration time for both MoE models and dense models. PP-out is more sensitive to latency for dense models than MoE models. This is because MoE layer has a longer computation time, which overlaps with cross-building P2P and effectively minimizes pipeline bubbles due to cross-building communication delays. Meanwhile, the shorter computation time of the dense model exposes such delays, leading to bubbles that slows down the iteration.

\begin{figure}[t]
    \centering
    \begin{subfigure}[h]{0.47\textwidth}
        \centering
        \includegraphics[width=\linewidth]{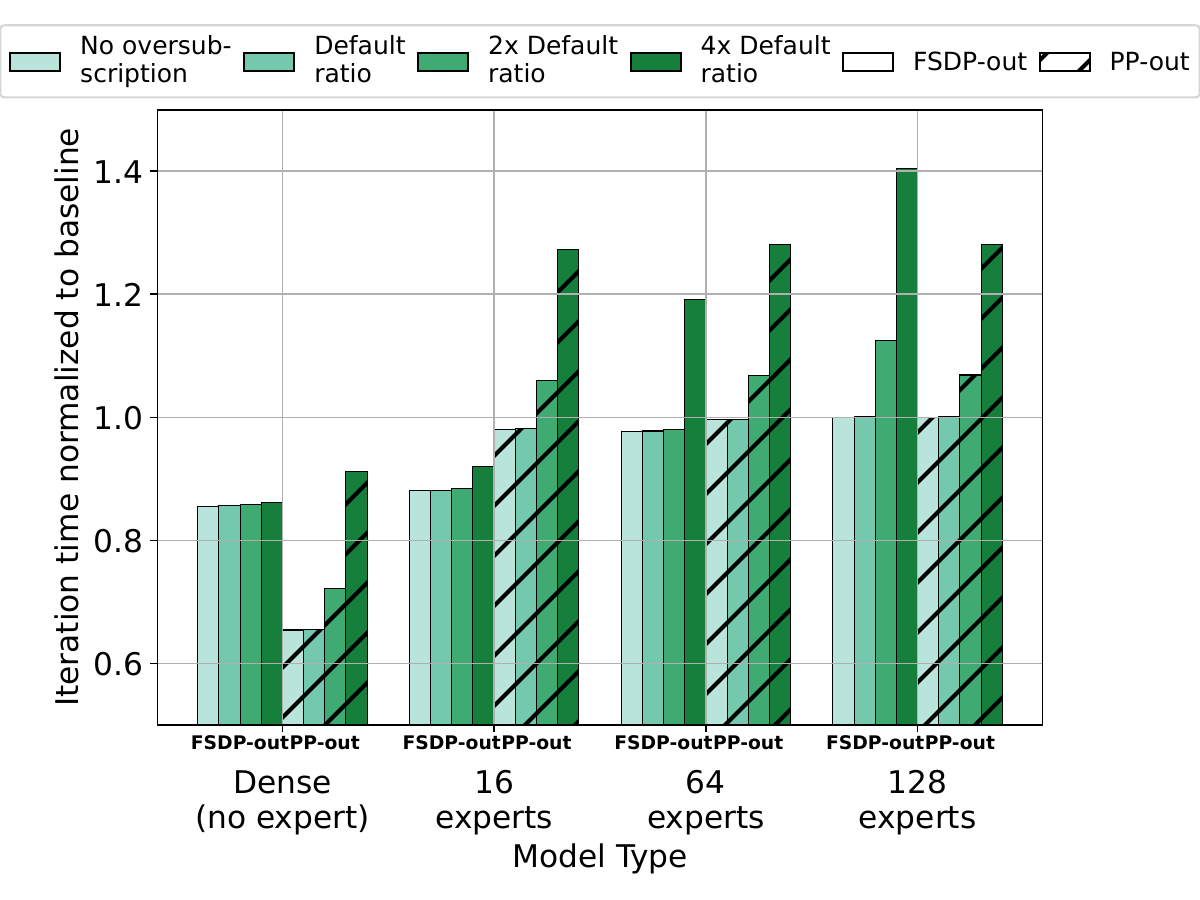}
        \caption{Varying bandwidth oversubscription ratios.}
        \label{fig:moe_bdw_dp1028}
    \end{subfigure}
    \begin{subfigure}[h]{0.47\textwidth}
        \centering
        \includegraphics[width=\linewidth]{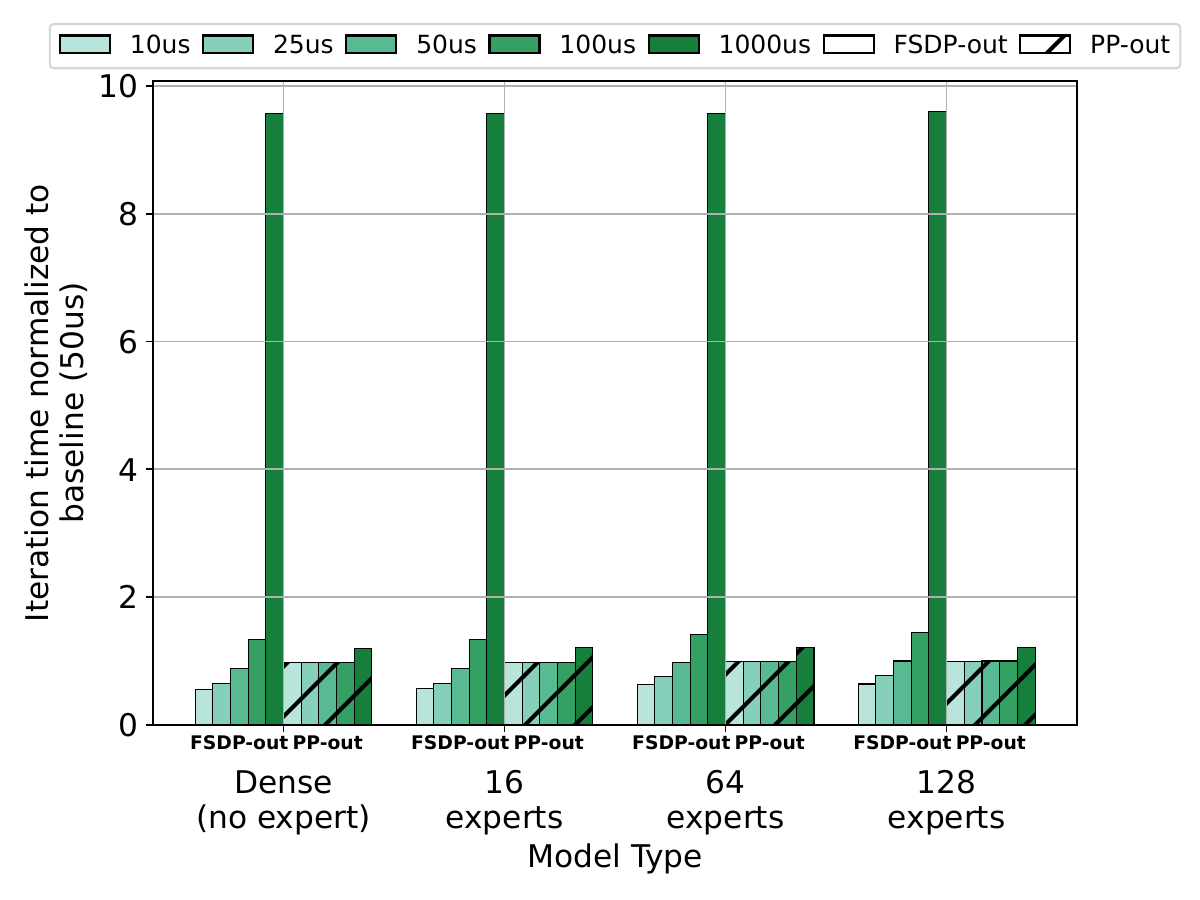}
        \caption{Varying link latencies.}
        \label{fig:moe_latency_dp1028}
    \end{subfigure}
    \caption{[Simulation] Training time comparison between MoE models with varying number of experts and the dense model, under parallelism configuration TP4-PP2-DP1024.}
    \label{fig:moe_vary_exps_dp1028}

\end{figure}

We configure TP degree as 4, PP degree as 2, and DP degree as 1024 for the second set of experiments. As shown in Figure~\ref{fig:moe_vary_exps_dp1028}, increasing oversubscription ratios across the datacenter building leads to substantial increase in iteration time for MoE models, but less so for the dense model training using FSDP-out. For dense and MoE models with fewer experts under constrained bandwidth (e.g. 4$\times$ default oversubscription ratio), FSDP-out has a slower iteration time increase than PP-out. This is due to a smaller communication volume with fewer experts and for the dense model. In contrast, for the 128 expert MoE model, FSDP-out starts seeing a faster increase in iteration time as the oversubscription ratio increases. Increasing latency across the datacenter building leads to increase in iteration time for both MoE models and dense models.
The increase is especially pronounced for FSDP-out at longer distances (e.g., 1000 us latency). This is caused by DP groups having 1024 GPUs, leading to each step in the ring transferring only a small chunk of data and link latency dominating the overall communication time. Additionally, the number of DP communication collectives is determined by the number of model chunks per pipeline stage. With just two stages and one layer per model chunk, each stage contains 31 chunks in total. This results in 62 DP collectives per iteration (31 parameter AllGather's and 31 gradient ReduceScatter's with FSDP), which accounts for a large fraction of the overall iteration time. In comparison, PP-out is less sensitive to long distance communication. 

\begin{boxB}
\textbf{Takeaways:} {PP-out is advantageous for training Mixture-of-Experts (MoE) models. Topology-aware model chunk sizing comes with throughput improvement potential.} 
\end{boxB}

\subsection{Topology Variants Impact}
\begin{figure}[t]
    \centering
    \begin{subfigure}[h]{0.47\textwidth}
        \centering
        \includegraphics[width=\linewidth]{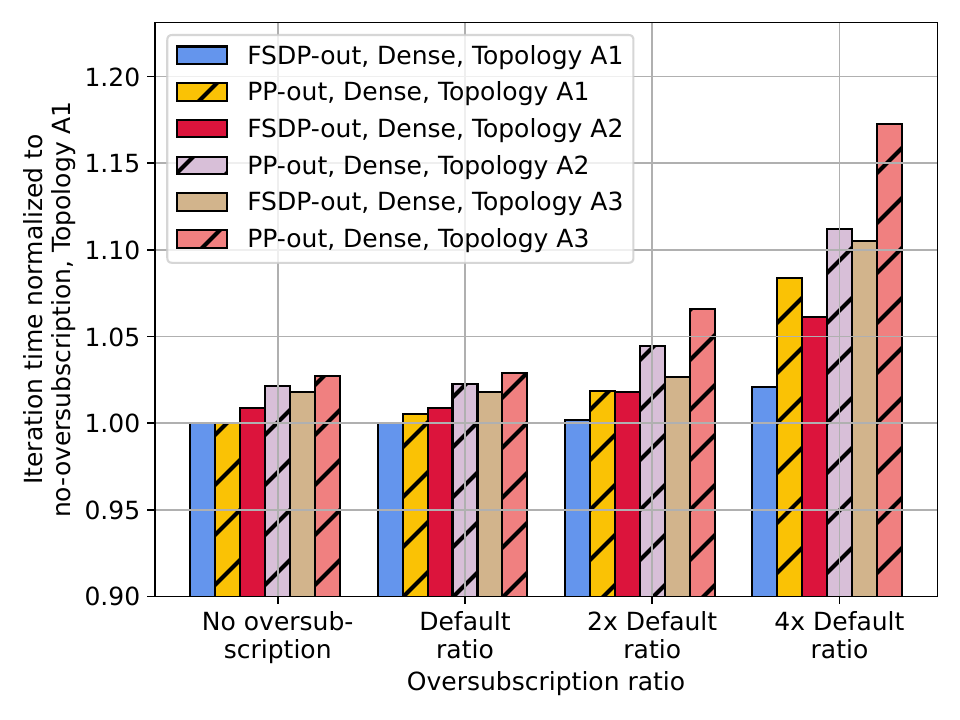}
        \caption{Varying cross-building oversubscription ratio with topology A1-A3.}
        \label{fig:nsf_intrabuilding_bdw}
    \end{subfigure}
    \begin{subfigure}[h]{0.47\textwidth}
        \centering
        \includegraphics[width=\linewidth]{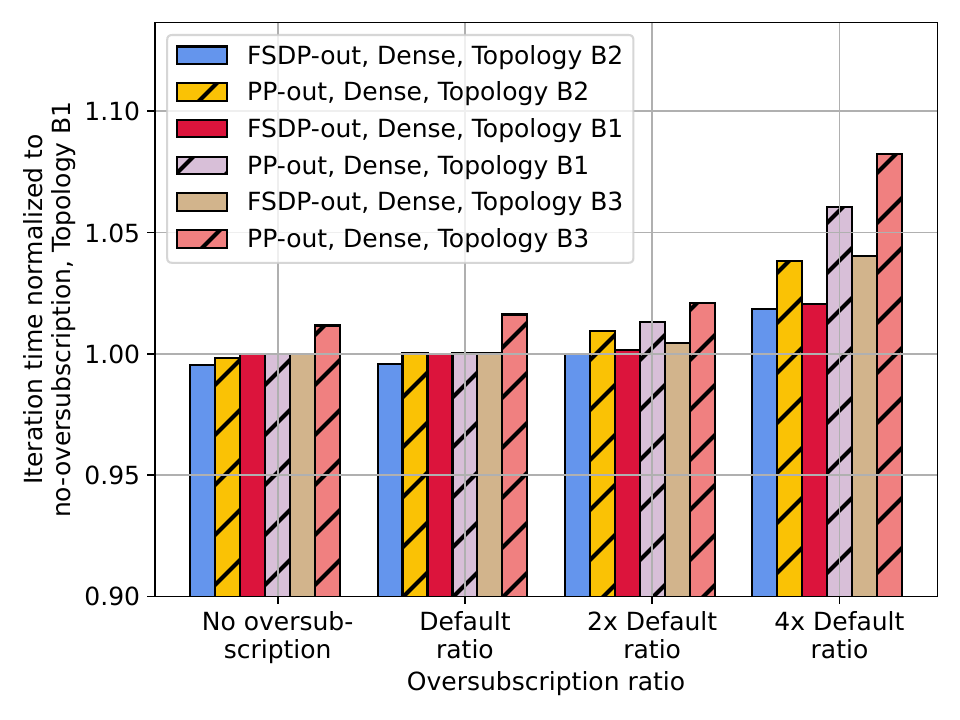}
        \caption{Varying cross-building oversubscription ratio with topology B1-B3.}
        \label{fig:aizone_intrabuilding_bdw}
    \end{subfigure}
    \caption{[Simulation] Impact of cross-building oversubscription on topologies with different intra-building oversubscription ratio. Iteration time normalized against the baseline scenario of A1 and B1 under their respective default cross-building oversubscription ratio. In the bottom figure, B2 (instead of B1) is placed on the left because B1 has a lower oversubscription ratio (1:1).}
    \label{fig:vary_intrabuidling_oversub_bdw}
\end{figure}

\begin{figure}[t!]
    \centering
    \begin{subfigure}[h]{0.47\textwidth}
        \centering
        \includegraphics[width=\linewidth]{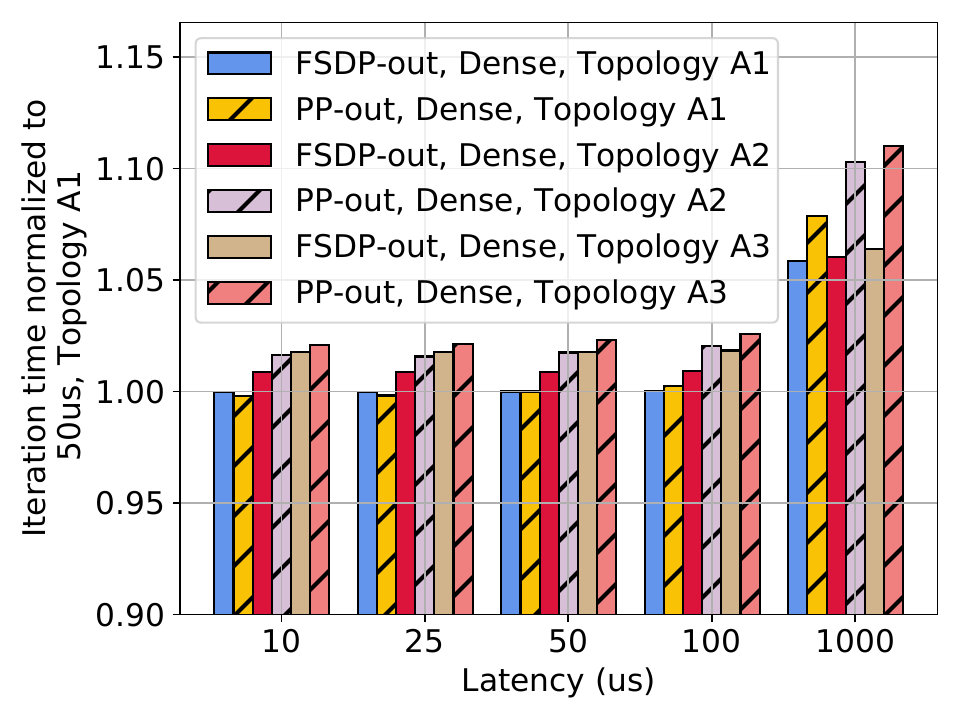}
        \caption{Varying cross-building latency with topology A1-A3.}
        \label{fig:nsf_intrabuilding_latency}
    \end{subfigure}
    \begin{subfigure}[h]{0.47\textwidth}
        \centering
        \includegraphics[width=\linewidth]{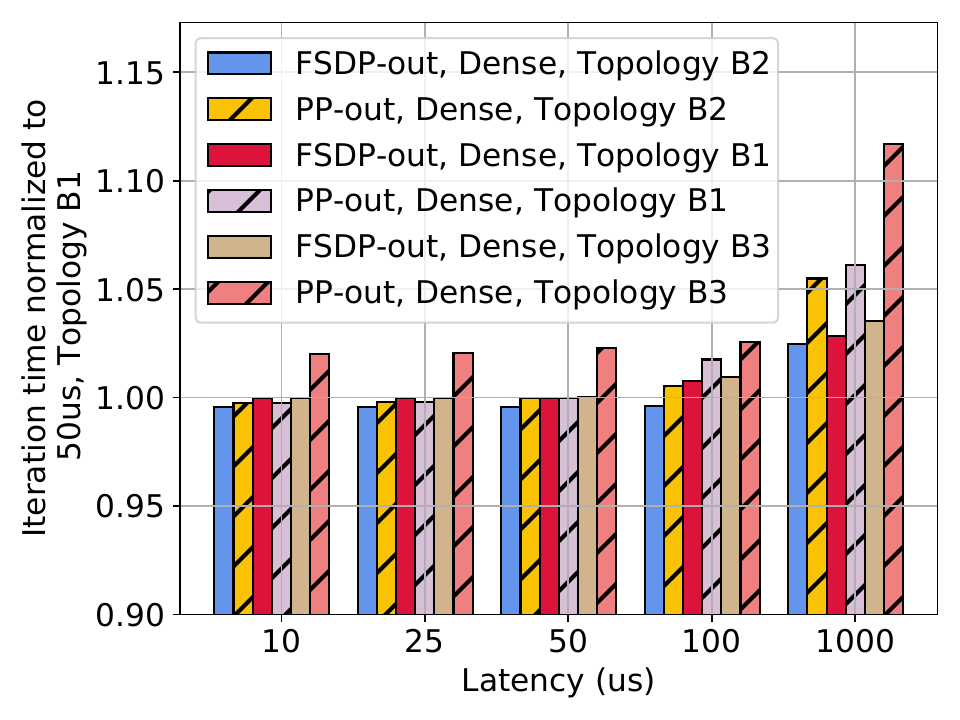}
        \caption{Varying cross-building latency with topology B1-B3.}
        \label{fig:aizone_intrabuilding_latency}
    \end{subfigure}
    \caption{[Simulation] Impact of cross-building latency on topologies with different intra-building oversubscription ratio. Iteration time normalized against the baseline scenario of A1 and B1 under 50~$\mu$s link latency. In the bottom figure, B2 (instead of B1) is placed on the left because B2 has a lower oversubscription ratio (1:1).}
    \label{fig:vary_intrabuidling_oversub_latency}
\end{figure}

To verify that the qualitative takeaways we derived from previous simulations will hold for other intra-building topologies, we simulate the 100K dense model workload on several topology variants. The detailed configuration of each variant is listed in Table~\ref{table:topology_configs}. In Figure~\ref{fig:vary_intrabuidling_oversub_bdw} and Figure~\ref{fig:vary_intrabuidling_oversub_latency}, we compare how iteration changes on topologies with different intra-building oversubscription ratios as we change the cross-building network conditions. PP-out iteration time grows faster as we increase the cross-building oversubscription ratio or latency. This trend is consistent with the dense model results in Section~\ref{simulation:pp_vs_dp}. When the intra-building oversubscription ratio is high, cross-building network has a more notable impact on iteration time, likely due to a longer queuing delay and more congestion events at the top layer switches of each building.

\begin{figure}[t]
    \centering
    \begin{subfigure}[h]{0.46\textwidth}
        \centering
        \includegraphics[width=\linewidth]{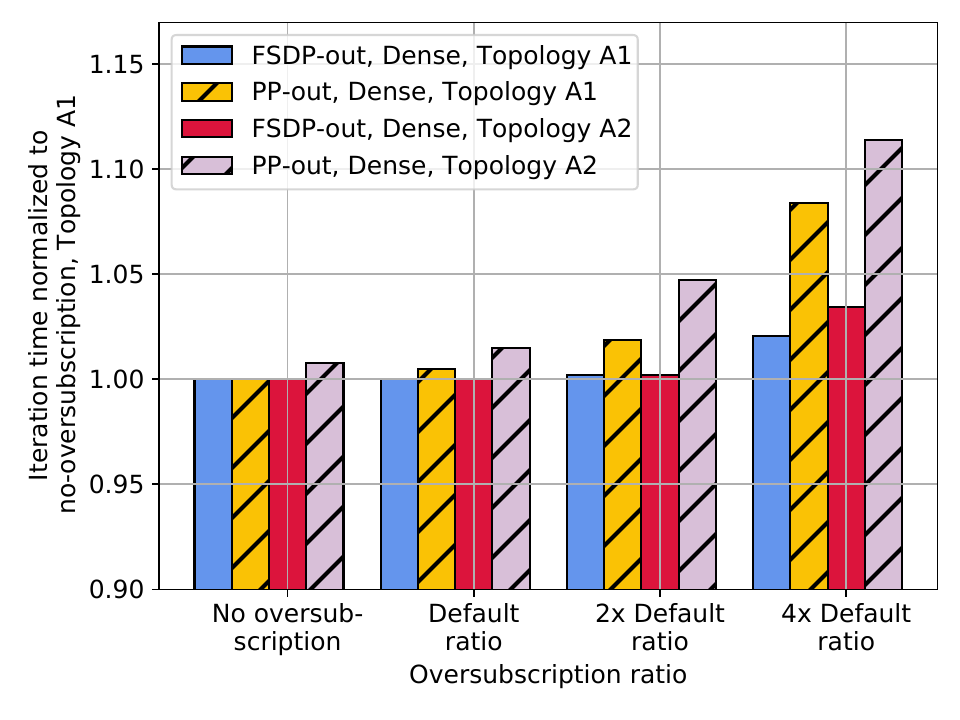}
        \caption{Varying cross-building oversubscription ratio with topology A1 and A4.}
        \label{fig:nsf_buildingcount_bdw}
    \end{subfigure}
    \begin{subfigure}[h]{0.46\textwidth}
        \centering
        \includegraphics[width=\linewidth]{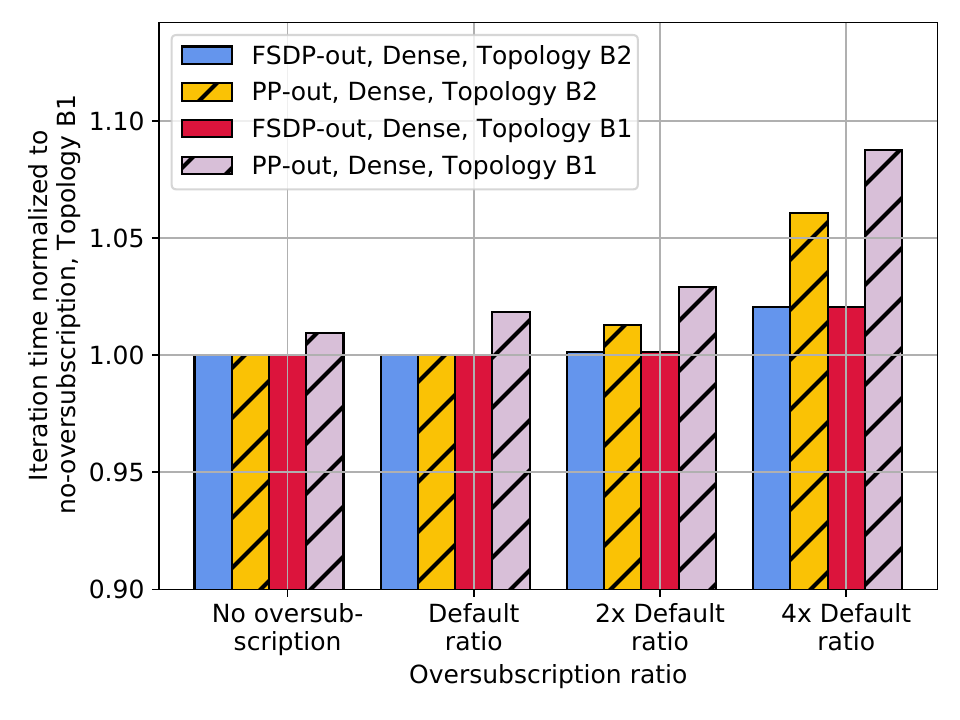}
        \caption{Varying cross-building oversubscription ratio with topology B1 and B4.}
        \label{fig:aizone_buildingcount_bdw}
    \end{subfigure}
    \caption{[Simulation] Impact of cross-building oversubscription on topologies with different number of GPUs per building. Iteration time normalized against the baseline scenario of A1 and B1 under their respective default cross-building oversubscription ratio.}
    \label{fig:vary_buidling_count_bdw}

\end{figure}

\begin{figure}[t!]
    \centering
    \begin{subfigure}[h]{0.46\textwidth}
        \centering
        \includegraphics[width=\linewidth]{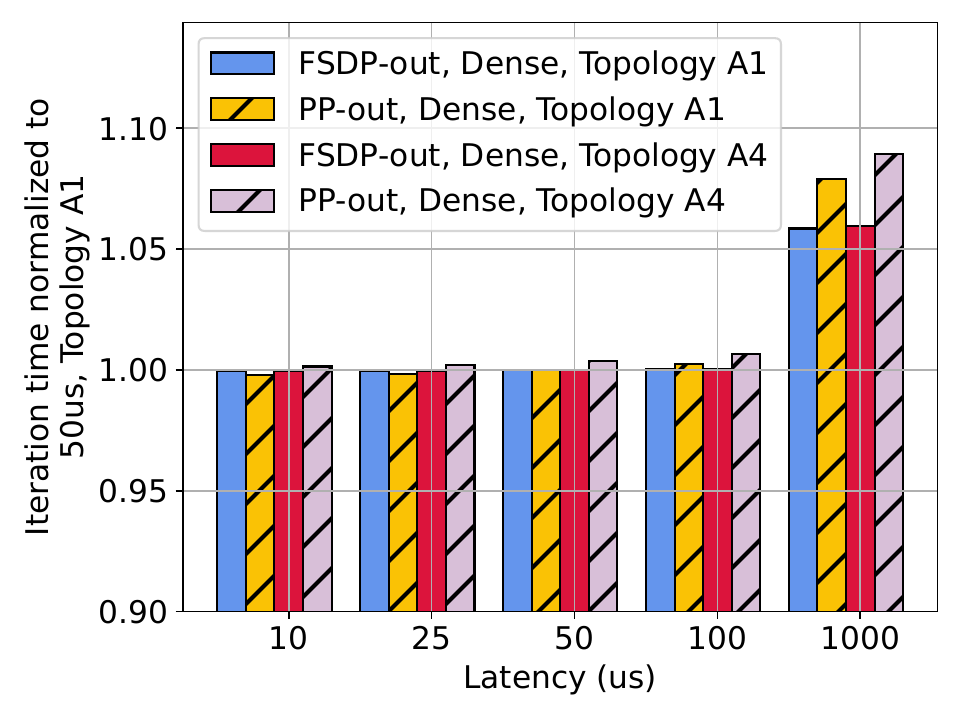}
        \caption{Varying cross-building latency with topology A1 and A4.}
        \label{fig:nsf_buildingcount_latency}
    \end{subfigure}
    \begin{subfigure}[h]{0.46\textwidth}
        \centering
        \includegraphics[width=\linewidth]{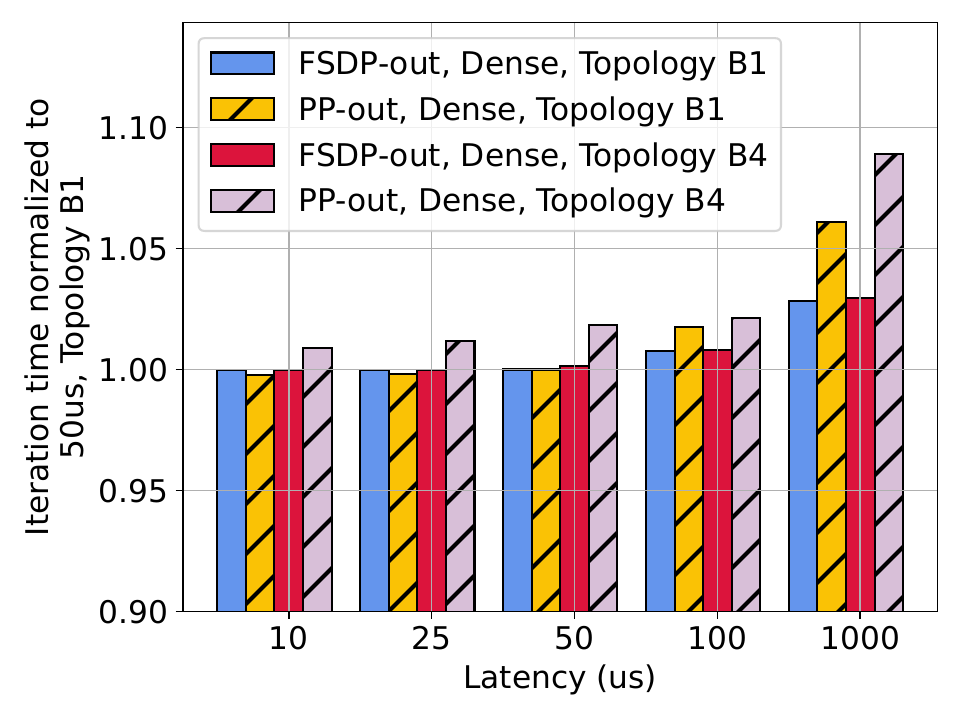}
        \caption{Varying cross-building latency with topology B1 and B4.}
        \label{fig:aizone_buildingcount_latency}
    \end{subfigure}
    \caption{[Simulation] Impact of cross-building latency on topologies with different number of GPUs per building. Iteration time normalized against the baseline scenario of A1 and B1 under 50~$\mu$s link latency.}
    \label{fig:vary_buidling_count_latency}

\end{figure}

We further simulate configurations where the GPU count per building is reduced, leading to more buildings involved in the training process. To ensure a fair comparison, we adjust the cross-building network layer so that the default cross-building oversubscription ratio remains consistent between A1 and A4, as well as between B1 and B4. As shown in Figure~\ref{fig:vary_buidling_count_bdw} and Figure~\ref{fig:vary_buidling_count_latency}, increasing the number of buildings has a greater impact on PP-out performance. This is because the number of point-to-point communications across datacenter buildings rises, due to more pairs of adjacent PP stages being distributed across different buildings.

\section{Parallelism Scheduling}\label{sec:parallelism_scheduling}

\subsection{DP Communication Pattern}
\label{eval:emulate_hsdp}

\begin{figure*}[t]
\begin{subfigure}{0.33\textwidth}
  \centering
  \includegraphics[width=\textwidth]{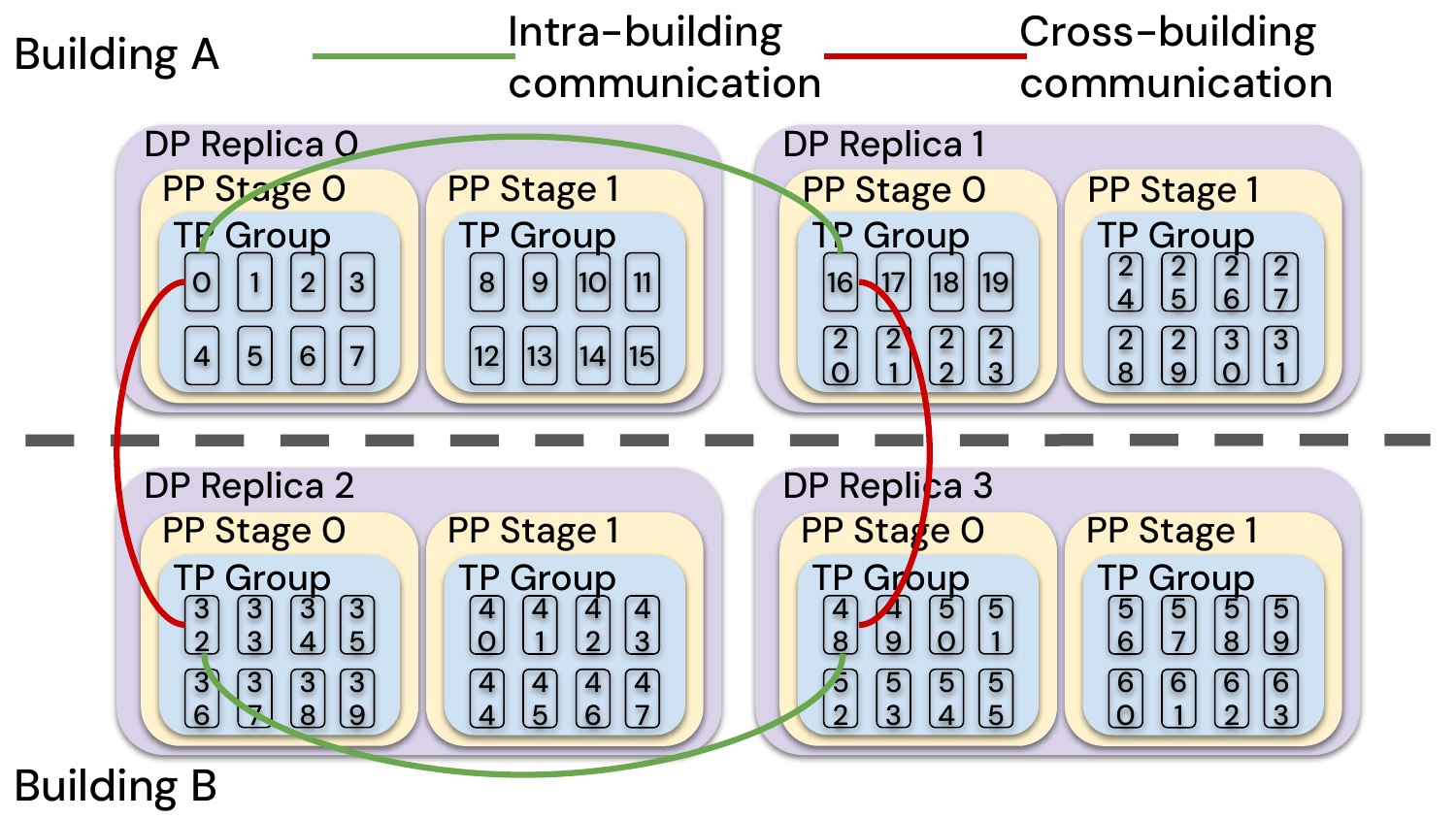}
  \caption{DP-out}
  \label{fig:dp_out_setting}
\end{subfigure}
\hfill
\begin{subfigure}{0.33\textwidth}
  \centering
  \includegraphics[width=\textwidth]{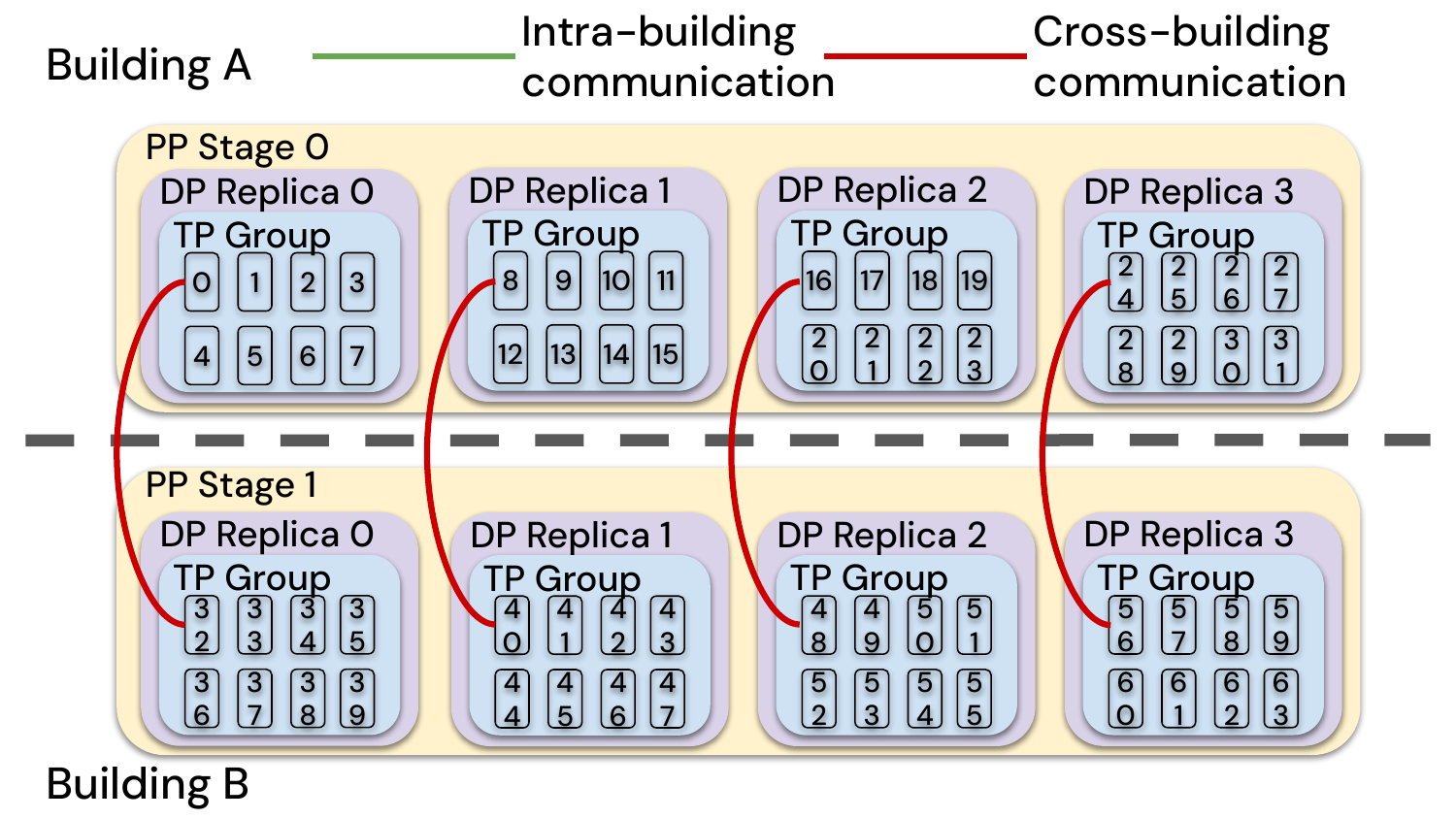}
  \caption{PP-out}
  \label{fig:pp_out_setting}
\end{subfigure}
\hfill
\begin{subfigure}{0.32\textwidth}
  \centering
  \includegraphics[width=\textwidth]{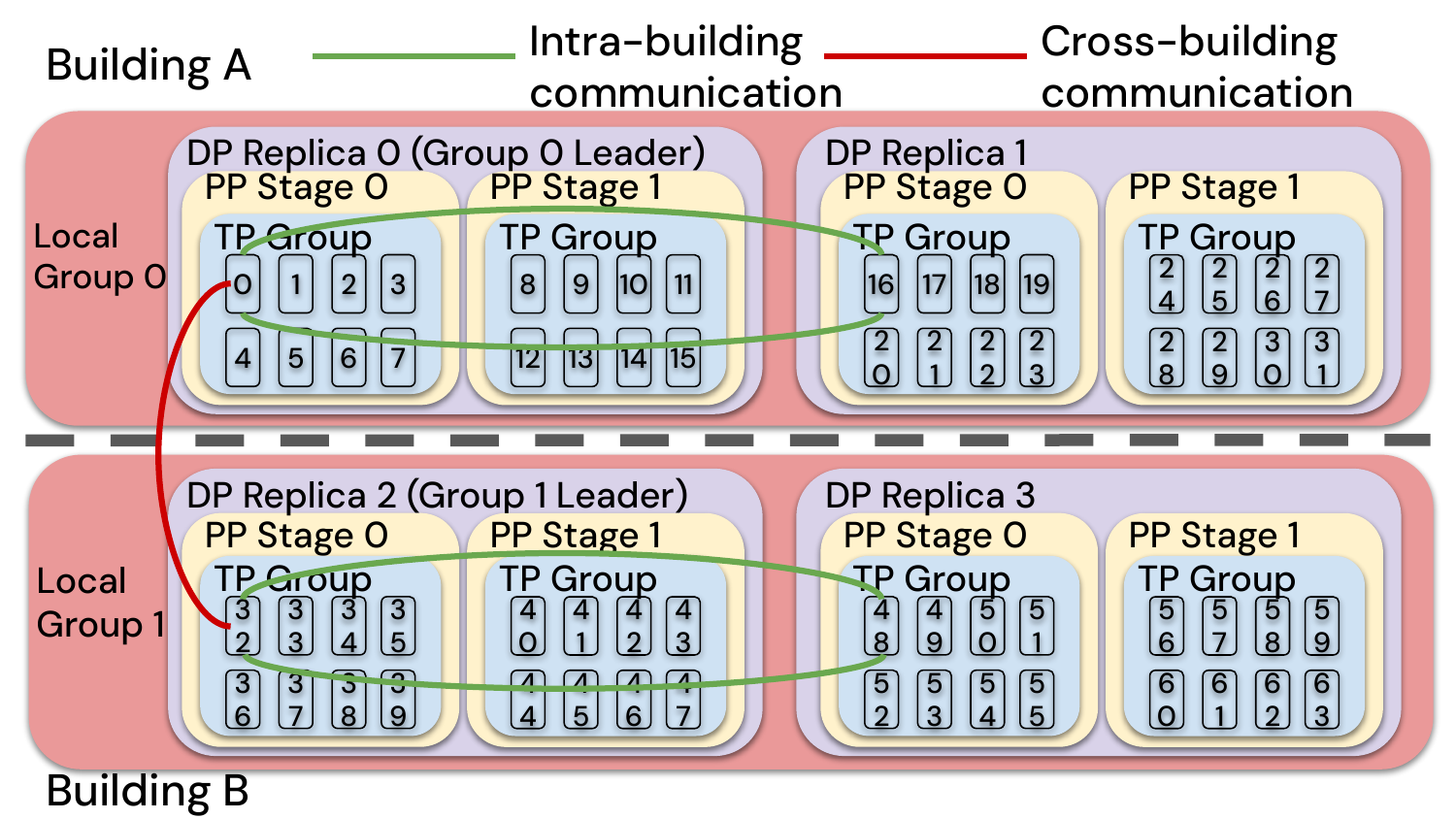}
  \caption{Hierarchical HSDP}
  \label{fig:hier_hsdp}
\end{subfigure}
\caption{Visualization of the emulation settings for DP-out, PP-out, and the hierarchical communication pattern of HSDP.}
\end{figure*}

Hierarchical DP has been adopted by LocalSGD works~\citep{localsgd, diloco} to enable varying synchronization frequency within and across groups. However, its communication pattern also works well for heterogeneous network links. Compared to FSDP, it reduces the total number of cross-building communication operations, as only the group leaders communicate over the cross-building links. However, this approach has a higher memory usage, as it has smaller FSDP groups and demands the group leaders to have a complete gradient during synchronization. This results in a lower maximum micro-batch size. We investigate the memory usage-throughput tradeoff of hierarchical DP by training the 17B dense model with the same setup as Section~\ref{emulaion:dp_pp_optimal_batch_size}, but employing a $2\times 2$ hierarchical HSDP, with Figure~\ref{fig:hier_hsdp} visualizing the DP replicas organization of this setting. DP replicas 0 and 1 form one local group, with replica 0 being the leader; DP replicas 2 and 3 form the other local group, with replica 2 being the leader. FSDP is used within local group, and DDP is used across local groups. Upon gradient synchronization, gradients are first aggregated within local groups, followed by one round of synchronization among the two group leaders.

\begin{figure*}[t]
    \begin{minipage}[b]{0.47\textwidth}
    \centering 
      \includegraphics[width=\linewidth]{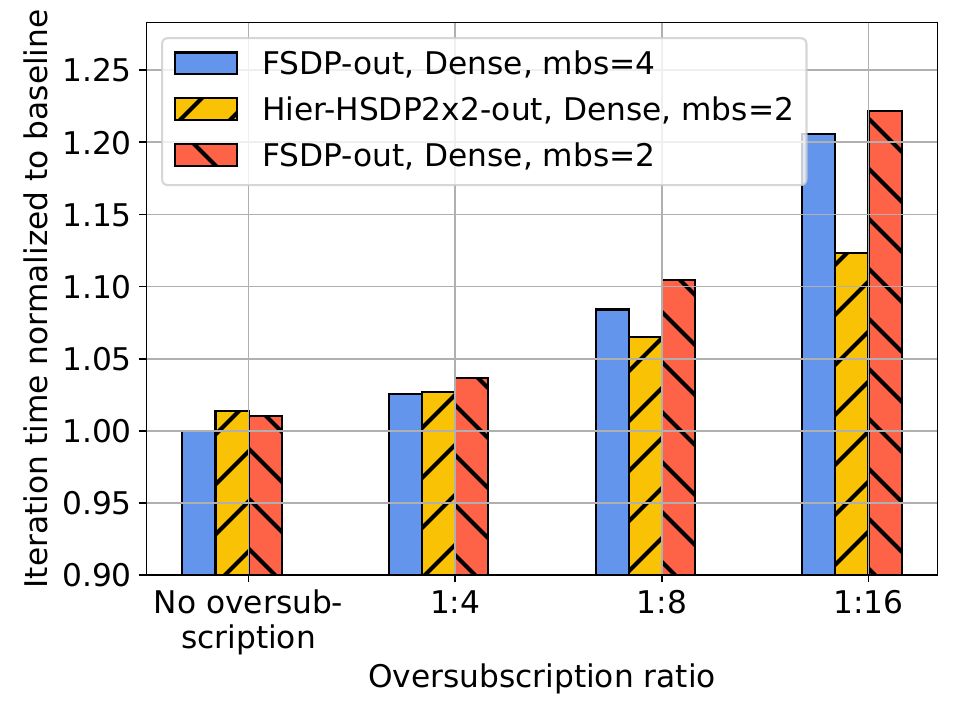}  
      \vspace{-0.25cm}
      \caption{\small [Testbed] Performance comparison between FSDP and HSDP. Iteration time are normalized to FSDP-out (microbatch size of 4). 
      }
      \label{fig:fsdp_hsdp_gbs176}
    \end{minipage}
    \hfill
    \begin{minipage}[b]{0.47\textwidth}
    \centering 
      \includegraphics[width=\linewidth]{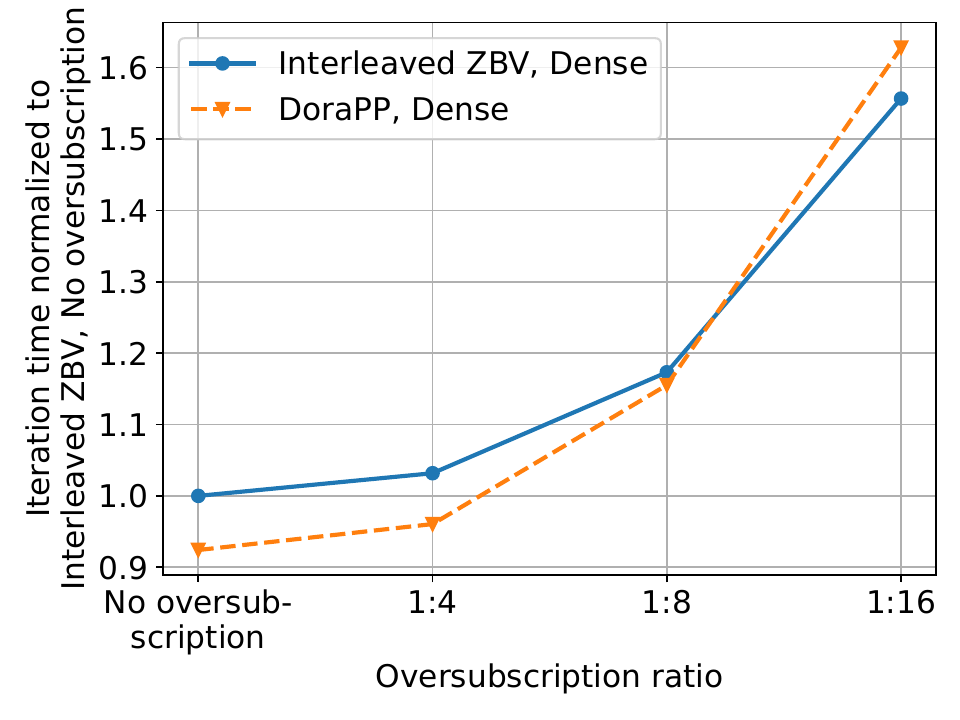}  
      \vspace{-0.25cm}
      \caption{\small [Testbed] PP-out iteration time with interleaved ZBV and DoraPP schedules. DoraPP increases cross-building communication frequency and is less effective under high oversubscription ratios.}
      \label{fig:pp_schedules} 
    \end{minipage}
    \vspace{-0.25cm}
\end{figure*}


As shown in Figure~\ref{fig:fsdp_hsdp_gbs176}, at moderate oversubscription, hierarchical HSDP performs slightly worse than FSDP due to GPU compute resource under-utilization and a worse communication-computation ratio caused by the smaller micro-batch size. However, at high oversubscription (e.g., beyond 1:4), HSDP's hierarchical aggregation design reduces cross-building communication frequency, leading to a 6.83\% iteration speedup at a 1:16 oversubscription ratio.

While hierarchical HSDP has a memory usage versus communication overhead tradeoff, the trend toward scale-across training is favorable to it. From a convergence perspective, the critical batch size for recent large language models is usually around tens of millions tokens~\citep{deepseekv3,bytedance_seed,moonshot_moonlight}; at tens to hundreds GPU scale with many pipeline stages, batch size constraints will likely require microbatch size as small as one. In such scenarios, HSDP’s hierarchical aggregation pattern can be leveraged without concern for compute under-utilization compared to FSDP.

\begin{boxB}
\textbf{Takeaways:} 
By reducing cross-building communication frequency, hierarchical aggregation offsets the performance loss typically associated with small microbatch sizes and improves training throughput under high oversubscription ratios. 
\end{boxB}

\subsection{Pipeline Schedules}
\label{sec:pp_schedule_model_chunk}

PP-specific optimizations focus on minimizing scheduling bubbles. However, the high communication cost over cross-building links bring in new tradeoffs to consider. As explained in Section~\ref{background:scale-across-prod-choices}, interleaved ZBV has an inferior performance than DoraPP in intra-building training due to its longer computation time. However, in scale-across setting, its V-shaped execution pattern leads to fewer cross-building P2P communication. 

To illustrate this, we train the 17B dense model (same setup as Section~\ref{emulaion:dp_pp_optimal_batch_size}) with PP-out. As shown in Figure~\ref{fig:pp_schedules}, DoraPP outperforms interleaved ZBV at lower oversubscription ratios thanks to its computational efficiency. However, its ``wrap-around'' pattern leads to more cross-boundary communication. At high oversubscription ratios where network is the bottleneck(e.g., beyond 1:8), minimizing traffic over the constrained links becomes important and interleaved ZBV becomes faster.

\begin{boxB}
\textbf{Takeaways:} 
Scale-across configurations benefit from bi-directional schedules -- that is, model chunks are assigned in the first-to-last-stage and then last-to-first order (e.g., interleaved ZBV).
\end{boxB}


\section{Network Layer Technologies}
\label{sec:transport_layer}
While it is intuitive that longer links have increased network latency and thus prolong the end-to-end training iteration time, their impact goes far beyond simple message delays. Long links often introduce reliability issues, and prolonged latency can undermine the effectiveness of load balancing and congestion control mechanisms. By simulating and analyzing these network conditions, we derive insights that inform transport layer design for scaling-across training.


\subsection{Collectives Performance at Long Distances}

\begin{figure}[t]
    \centering
    \includegraphics[width=0.93\textwidth]{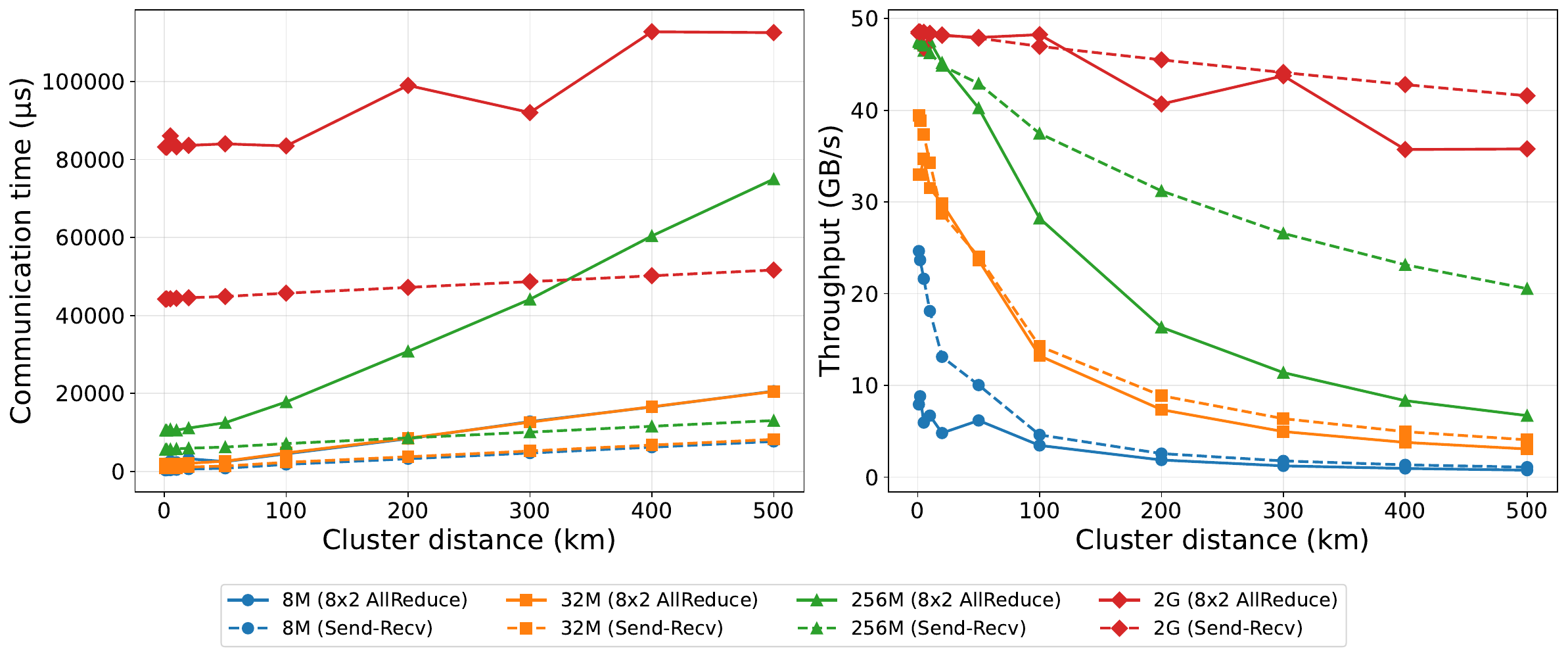}
    \caption{\small Completion time and throughput of NCCL Send-Recv between a pair of GPUs AllReduce among 16 GPUs located in two clusters at varying distances.}
    \label{fig:ar8x2_sedrecv_at_distances}
    \vspace{-0.2cm}
\end{figure}

We study NCCL collective completion times at varying distances with GPUs located in two clusters. Overall, we observe that the completion time increases linearly with cross-cluster distance, except for the 2GB AllReduce operation among 16 GPUs (eight per cluster). This linear trend supports our approach of simulating longer inter-cluster distances by increasing link latency and guides the communication time modeling of \sysname (Section~\ref{optimization}).

For a given message size, the maximum achievable throughput is $\frac{\text{Message Size}}{\text{Latency}}$. Consequently, larger message sizes are more effective at approaching line rate over long distances. We also observe that point-to-point Send-Recv outperforms AllReduce. This is because AllReduce, being a more complex collective operation, incurs higher processing stack overhead and requires careful tuning of communication buffers and channels to maintain high throughput. Optimizing AllReduce for large message sizes is challenging, as the available memory for communication buffers without affecting application performance becomes limited. With a 2GB message size, AllReduce requires manual tuning at distances longer than 100 km, leading to the non-monotonic completion time curve.

\subsection{Latency Impact}
\label{subsec:simulation_latency}
\begin{figure}[t]
    \centering
    \begin{subfigure}[h]{0.47\textwidth}
        \centering
        \includegraphics[width=\linewidth]{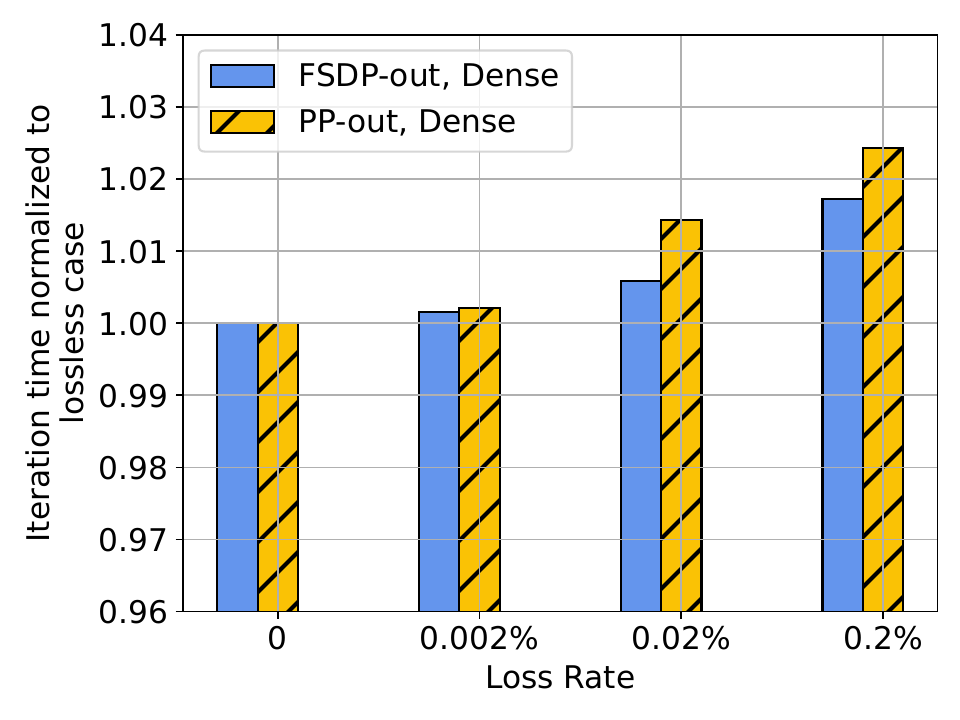}
        \caption{\small Impact of increasing packet loss rate with a 50 $\mu$s link latency.}
        \label{fig:packet_loss_impact}
    \end{subfigure}
    \begin{subfigure}[h]{0.47\textwidth}
        \centering
        \includegraphics[width=\linewidth]{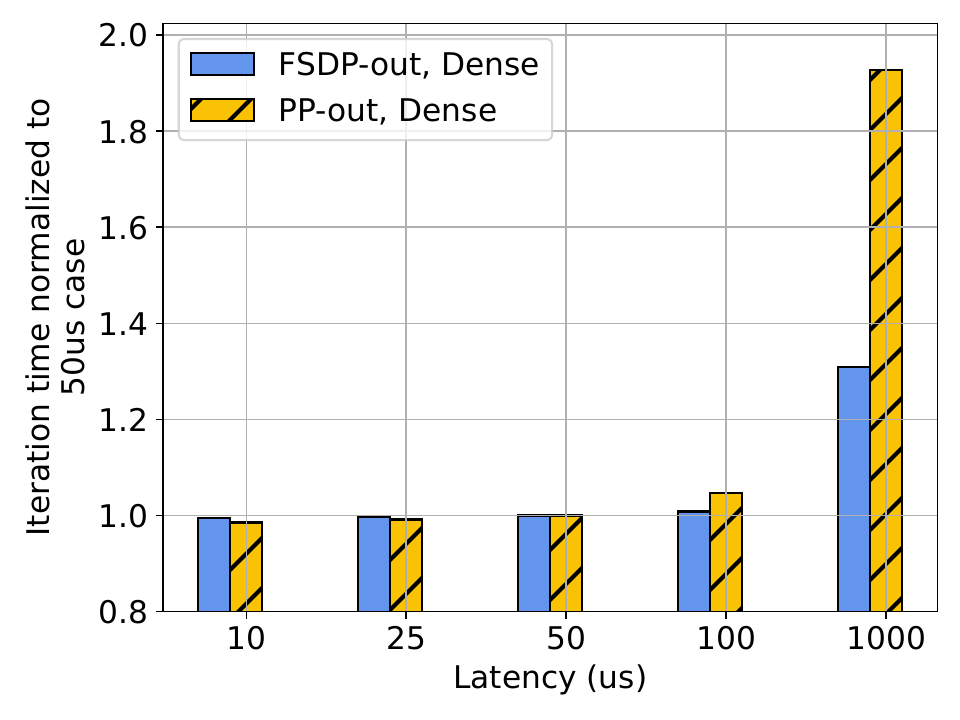}
        \caption{\small Impact of increasing link latency at a fixed packet loss rate of 0.02\%.}
        \label{fig:packet_loss_latency_impact}
    \end{subfigure}
    \vspace{-0.2cm}
    \caption{\small [Simulation] PP-out is more sensitive to packet loss. Higher link latencies substantially increase the cost of lost packet retransmissions, resulting in slower training.}
    \label{fig:large_scale_packet_loss_impact}

\end{figure}
\parab{Reliability.}
Reliability challenges become more prominent as network distance grows. With greater distances, both packet loss rates and latency variation tend to rise. Our observations indicate that cross-region packet loss rates in our data centers is more than 20× higher than that of intra-building communication. In geo-distributed environments, these issues are even more pronounced: loss rates can range from 0.1\% to 10\%, and the 99th percentile latency may be one to two orders of magnitude higher than the median latency in such settings~\citep{msft_sdr,deconstructing_tail_at_scale,tail_at_scale}.

We investigate the impact of packet loss on PP-out and DP-out performance using simulation. For loss recovery, we employ Go-Back-N~\citep{rdma_gobackn}, which is the default protocol for RDMA~\citep{fec_for_rdma,rdma_congestion}. To isolate the effects of increased loss rates and longer latencies, we conduct two sets of experiments: (1) fix latency at 50 microseconds and vary the packet loss rate from 0.002\% to 0.2\%, and (2) fix the loss rate at 0.02\% and vary latency from 10 to 1000 microseconds.
As shown in Figure~\ref{fig:packet_loss_impact} and Figure~\ref{fig:packet_loss_latency_impact}, PP-out exhibits greater sensitivity to increased packet loss rates. This sensitivity arises from pipeline bubbles caused by timeouts and retransmissions during each point-to-point activation and gradient communication. In contrast, DP-out is less affected, as its communication frequency across long-distance links is low, and only the parameter all-gather and gradient reduce-scatter operations of the first layer are on the critical path.

In Figure~\ref{fig:packet_loss_latency_impact}, we observe that at a link latency of 1000~$\mu$s, the iteration time increases by 1.30$\times$ for DP-out and 1.93$\times$ for PP-out compared to the baseline case of 50~$\mu$s link latency. These increases are substantially larger than those observed without packet loss in Figure~\ref{fig:nsf_intrabuilding_latency} (1.07$\times$ and 1.10$\times$ for DP-out and PP-out respectively on Topology A1). This gap highlight that latency's impact on training iteration time is amplified when the network is lossy, as the packet loss recovery cost is primarily determined by the Round-Trip-Time (RTT). 

It is also worth noting that the above analysis extends to latency variation, which tends to worsen over long distances. In such scenarios, tail latency is often a magnitude greater than the RTT and has a similar impact on training iteration time as the overhead from packet loss recovery.

\parab{Heterogeneous latency.}
\begin{figure}[t]
    \centering 
      \includegraphics[width=0.5\textwidth]{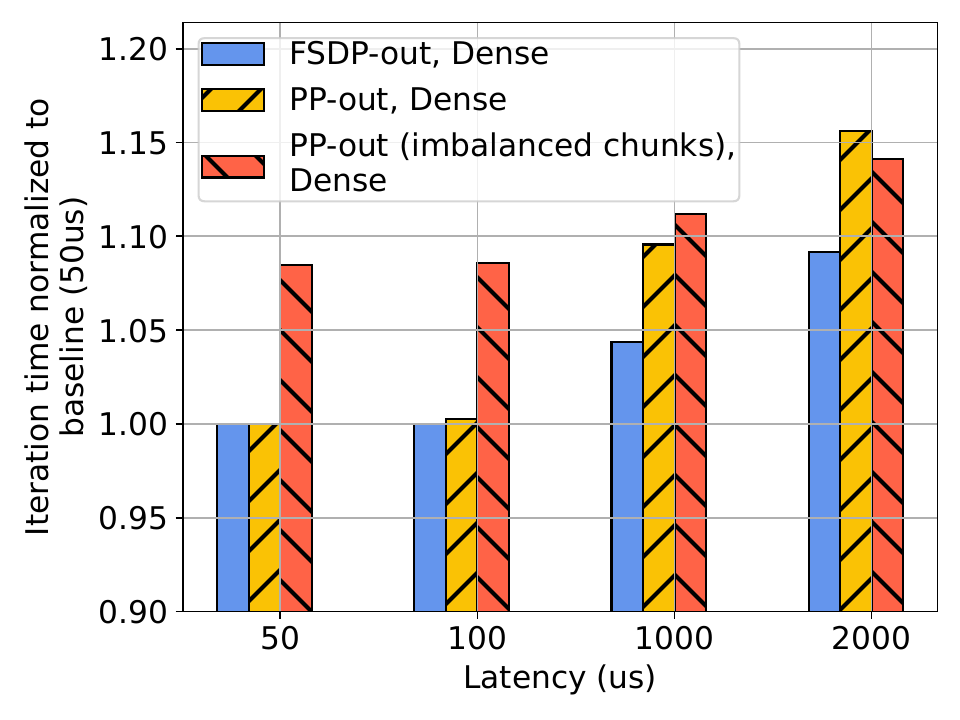} 
      \vspace{-0.2cm}
      \caption{\small [Simulation] Iteration times normalized against default case with a 50 $\mu$s link latency. Impact of having heterogeneous link latency due to varying distances between regions.}
      \label{fig:hetero_topo} 
      \vspace{-0.2cm}
\end{figure}

\begin{figure}[t!]
\centering
\includegraphics[width=0.75\linewidth,trim={0.4cm, 0cm, 0cm, 0cm}, clip]{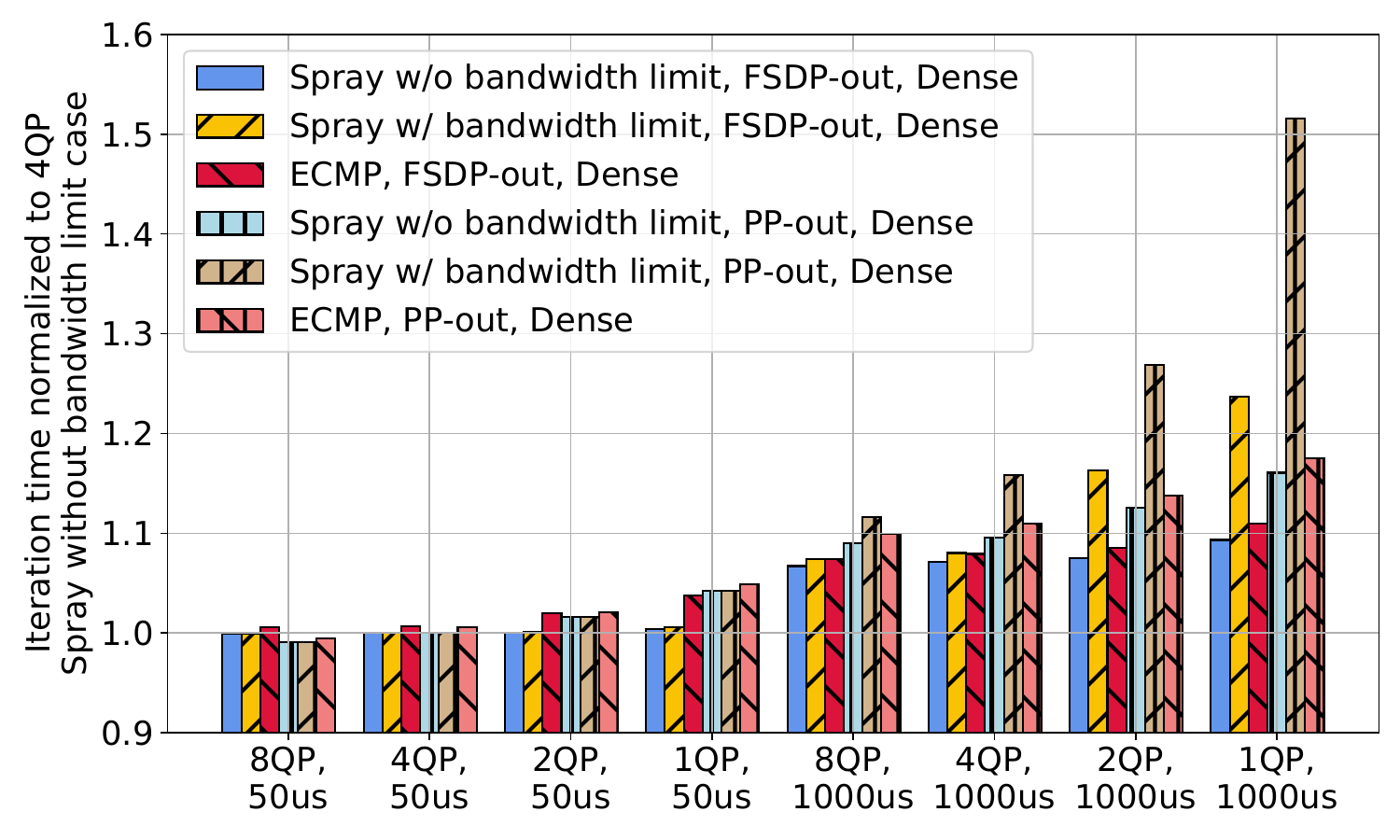}
\vspace{-0.25cm}
\captionof{figure}{\small [Simulation] Comparison of ECMP and Packet Spraying results with various number of Queue Pairs (QP). 
Realistic packet spraying that accounts for in-flight data limitation struggles under long latencies.
}
\vspace{-0.25cm}
\label{fig:ecmp_spraying}
\end{figure}

We also model a heterogeneous topology scenario where buildings locate in different regions. We set the default inter-building latency as 50$\mu$s, but pick one building as a distanced building and vary the latency between this building and other buildings between 50 to 2000 $\mu$s. Besides comparing PP-out and DP-out, we also enable imbalanced model chunk size for PP-out so that one model chunk has twice as many layers as the other model chunks. Imbalanced chunk size reduces the total number of model chunks (and hence the frequency of cross-region communication) and introduces more FLOPs for the larger chunk, which can improve computation-communication ratio to mask long latency, but can also become the bottleneck when the latency is not long enough. As shown in Figure~\ref{fig:hetero_topo}, enabling imbalanced chunk size makes PP-out slower initially. However, when the latency reaches 2000 $\mu$s, imbalanced chunk size leads to a 1.3\% iteration speed up over PP-out.

\begin{boxB}
\textbf{Takeaways:} For training over ultra long distances (e.g., beyond 20 kilometers with longer than 100 microseconds latency), proactive measures such as selective redundancy and erasure encoding are worth considering.  
When cross-region latency is high, reducing the frequency of cross-region communication becomes more important for PP-out---this favors lower interleaving (fewer model chunks) by increasing model chunk size.
\end{boxB}

\subsection{ECMP vs. Packet Spraying}
In this subsection, we present controlled simulation results to characterize the throughput of load balancing techniques under long latencies.
Packet spraying~\citep{packet_spray} is generally believed to outperform the hash-based Equal-Cost Multi-Path (ECMP)~\citep{ecmp} routing~\citep{impact_packet_spray, case_for_packet_spray, enabling_packet_spray}. However, the throughput of packet spraying is limited by the \newchange{overhead of metadata to track in-flight packets on a per Queue Pair (QP) basis} at the network interface controller (NIC).
Between each QP, running packet spraying leads to out-of-order packet arrival, as packets might take different paths with varying latencies. \newchange{To handle out-of-order delivery and support reliable transport, NICs must track additional per-packet metadata for in-flight packets on each QP. This bookkeeping overhead imposes an upper bound on the number of outstanding packets that can be in flight per QP.}
NVIDIA ConnectX-6 sets this upper bound as \newchange{$2^9=512$ (the maximum \text{log\_tx\_psn\_win} value is 9)}~\citep{nvidia_connectx6dx}, which leads to a maximum of 2MB in-flight data when using a 4KB packet payload. When link latency is 1000 $\mu$s, per QP throughput drops to $\frac{\text{2MB}}{1000\mu\text{s}} = 16$ Gbps with packet spraying. 
In this section, we compare packet spraying and ECMP at long distances, both with and without accounting for the in-flight packets limitation.

We fix the oversubscription ratio at the default value and evaluate with the default latency of 50 $\mu$s as well as the more extreme case with latencies of 1000 $\mu$s. We vary the number of QPs per GPU pair between one to eight. When enforcing the bandwidth restriction for packet spraying, we keep the same number of cross-building links and directly change their bandwidth.
In Figure~\ref{fig:ecmp_spraying}, packet spraying outperforms ECMP with or without bandwidth limitation consideration at 50$\mu$s latency. With more than two QPs, the iteration time difference between simulating the bandwidth restriction or not is negligible. However, when link latency reaches 1000~$\mu$s and reduces the packet spraying's transmission speed upper-bound to merely 16 Gbps per QP, packet spraying becomes 0.10\% to 11.44\% slower than ECMP with no more than 4 QPs.

\begin{boxB}
\textbf{Takeaways:} Packet spraying is better for cross-building training due to avoiding the per-flow hashing that leads to persistent hash collisions and significant load imbalance under low-entropy training traffic. 
In contrast, for cross-region deployments, ECMP can be more effective because packet spraying throughput becomes limited by constraints on \newchange{the number of in-flight packets}. Consequently, when attempting to extend packet spraying to cross-region scenarios, increasing the number of QPs without NIC and switch resource exhaustion is required to sustain throughput.
\end{boxB}

\subsection{Congestion Control}
\begin{figure}[t]
\centering
\includegraphics[width=0.68\linewidth,trim={0.3cm, 0cm, 0cm, 0cm}, clip]{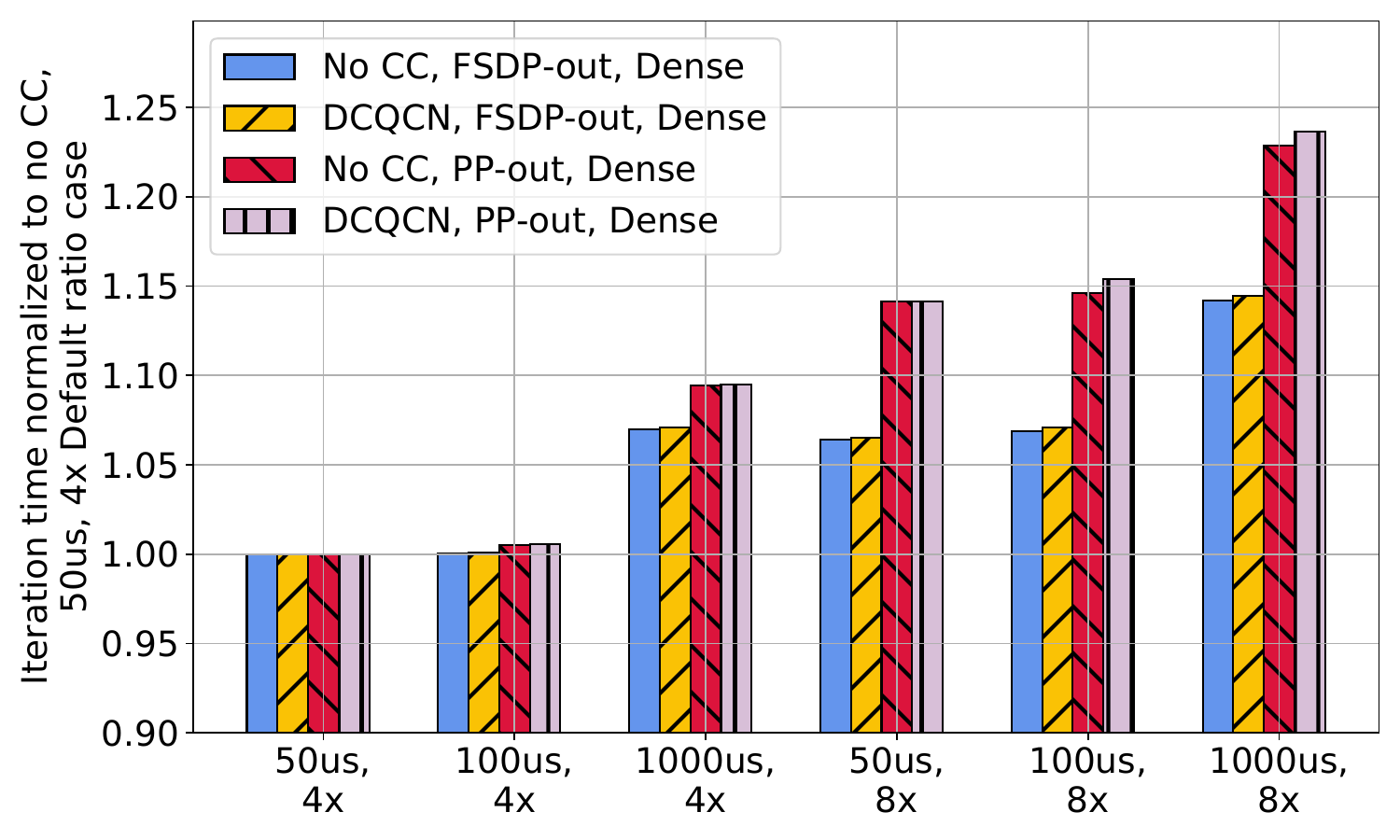}
\vspace{-0.2cm}
\captionof{figure}{[Simulation] Comparison of congestion control strategies. 
Under long link latencies and higher oversubscription ratio, not enabling congestion control is better than using an ECN based congestion control.}
\label{fig:congestion_control}
\vspace{-0.25cm}
\end{figure}

Explicit Congestion Notification (ECN~\citep{ecn})-based congestion control (CC) is widely adopted in modern data center. However, concerns arise regarding the effectiveness of ECN-based CC over high-latency links. ECN-based algorithms rely on feedback from the network, and the time required for endpoints to receive and react to ECN signals grows with link latency, making the control loop less responsive. To simulate congestion, we study the scenarios with a high oversubscription ratios that are 4$\times$ and 8$\times$ of the default ratio. We evaluate with the default latency of 50 $\mu$s as well as more extreme cases with latencies of 100 and 1000 $\mu$s.

At an oversubscription 4$\times$ the default ratio, disabling congestion control (CC) does not significantly affect end-to-end iteration time, regardless of link latency. This suggests that the network is not experiencing congestion under these conditions.  At an oversubscription 8$\times$ the default ratio, disabling CC still has minimal impact at a link latency of 50$\mu$s, indicating that ECN-based CC continues to perform well in typical cross-building scenarios. However, as we extend the latency to 100$\mu$s and 1000$\mu$s, disabling CC results in a modest training iteration speedup of 0.89\% and 1.52\% respectively.

While our simulations demonstrate only marginal gains, it is worth noting that these results are based on a single-tenant environment with only one training workload. In real-world deployments where multiple jobs are collocated, network congestion dynamics become significantly more complex. The presence of concurrent workloads can lead to higher buffer occupancy, greater variability in queue lengths, and more frequent congestion events. Under these conditions, the performance differences between various CC strategies are expected to be more pronounced.
\begin{boxB}
\textbf{Takeaways:} Disabling CC at ultra-long distances yields slightly better performance compared to applying DCQCN universally across all distances.
\end{boxB}

\section{{ScaleAcross} Explorer}
\label{optimization}
\begin{figure*}[t]
\centering
\includegraphics[width=0.97\linewidth,trim={0, 4cm, 10cm, 4.2cm}, clip]{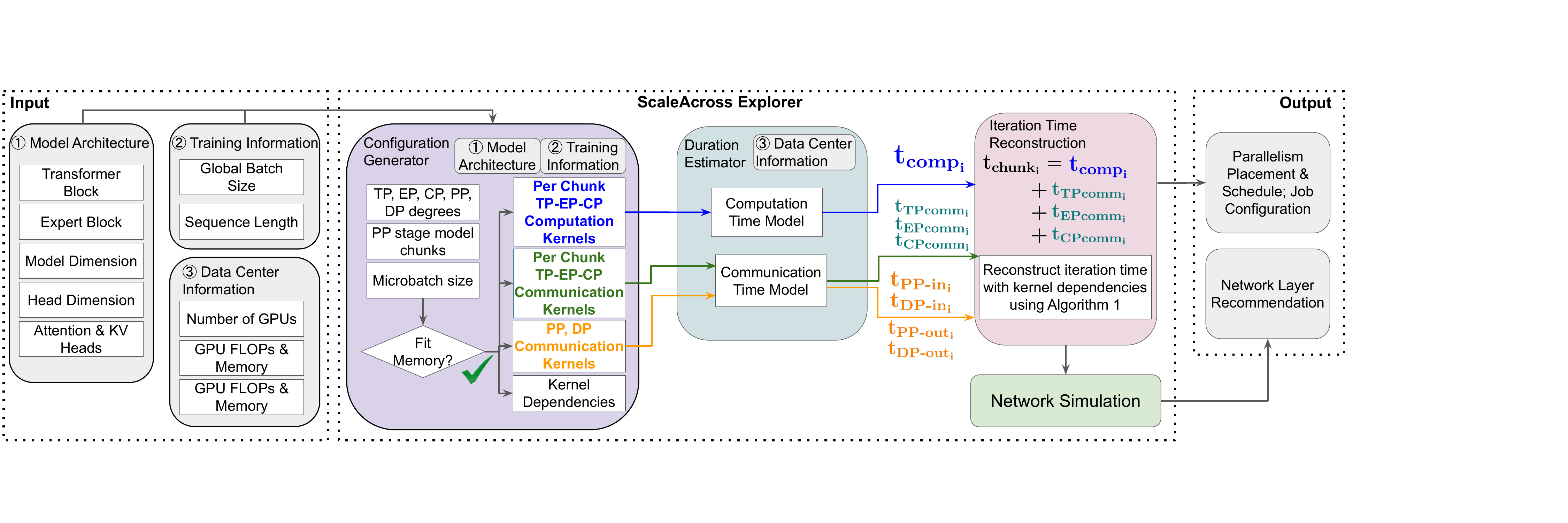}
\caption{\small \sysname workflow overview.}
\label{fig:optimizer_workflow}
\end{figure*}
Section~\ref{sec:parallelism_schemes} to Section~\ref{sec:transport_layer} characterizes individual design space dimensions. As it's challenging to capture their combined impact with a decision tree, we introduce \sysname, a configuration optimizer that given model architecture, batch size, and cluster settings, holistically optimizes microbatch size and parallelism configurations and recommends appropriate network transport layer mechanisms.

\begin{algorithm}[t]
\caption{End-to-end training iteration time reconstruction}
\label{alg:iteration_reconstruct}
\begin{algorithmic}
\Input
\Desc{$M$}{Model chunk TP-EP-CP kernels;}
\Desc{$P$}{PP communication kernels;}
\Desc{$D$}{DP communication kernels;}
\Desc{$E$}{$E[K]:=$ List of kernels that $K$ depends on;}
\EndInput
$T_\text{DPout} \gets \{\}$; $T_\text{PPout} \gets \{\}$;
$t_\text{DPout} \gets 0$;
$t_\text{PPout} \gets 0$;
\While{$M \neq \emptyset$ or $P \neq \emptyset$ or $D \neq \emptyset$}
\ForEach{$K_i$ whose parent kernels all finished}
\State \Comment{Find the latest finish time of parent kernels}
\State $t_\text{DPout\_start}=\max_{J \in E[K_i]} T_\text{DPout}[J]$;
\State $t_\text{PPout\_start}=\max_{J \in E[K_i]} T_\text{PPout}[J]$;
\If{$K_i \in M$}
    \State $T_\text{DPout}[K_i]=t_\text{DPout\_start} + \text{t}_{\text{chunk}_\text{i}}$;
    \State $T_\text{PPout}[K_i]=t_\text{PPout\_start} + \text{t}_{\text{chunk}_\text{i}}$;
    \State $M$.pop($K_i$);
\ElsIf{$K_i \in P$}
    \State $T_\text{DPout}[K_i]=t_\text{DPout\_start} + \text{t}_{\text{PP-in}_\text{i}}$;
    \State $T_\text{PPout}[K_i]=t_\text{PPout\_start} + \text{t}_{\text{PP-out}_\text{i}}$;
    \State $P$.pop($K_i$);
\Else
    \State $T_\text{DPout}[K_i]=t_\text{DPout\_start} + \text{t}_{\text{DP-out}_\text{i}}$;
    \State $T_\text{PPout}[K_i]=t_\text{PPout\_start} + \text{t}_{\text{DP-in}_\text{i}}$;
    \State $D$.pop($K_i$);
\EndIf
\State \Comment{Update DP-out and PP-out iteration end time}
\State $t_\text{DPout}=\max(t_\text{DPout}, T_\text{DPout}[K_i])$;
\State $t_\text{PPout}=\max(t_\text{PPout}, T_\text{PPout}[K_i])$;
\EndFor
\EndWhile
\State \Return $\min(t_\text{DPout}, t_\text{PPout})$;
\end{algorithmic}
\end{algorithm}

\subsection{Design Overview}
\sysname aims to minimize training iteration time for scaling-across training jobs spanning across diverse network characteristics. The end-to-end workflow is presented in Figure~\ref{fig:optimizer_workflow}. \sysname takes (1) model architecture (e.g., \# of layers, model dimension, FFN dimension, \# of experts, \# of attention and kv heads, etc.), (2) batch size, and (3) cluster settings (\# of accelerators and their specifications and network topology) as input, and solve for near optimal workload configurations in three steps:

\begin{enumerate}
    \item Parallelism configuration and placement: Generate all valid combinations of microbatch size, degrees of TP, EP, CP, PP, and DP, and model chunk sizes that fit the accelerator memory. And then generate communication groups for both DP and PP as the outermost and runs on the across network. 
    \item Parallelism scheduling: Estimate the computation and communication kernel time of each valid combination on the input topology, reconstruct the end-to-end iteration time with interleaved ZBV and DoraPP schedules using  Algorithm~\ref{alg:iteration_reconstruct}, and return the configuration with the fastest iteration.
    \item Network technology choices: Based on the configure returned by the previous step and the input topology, recommend load balancing, congestion control, and packet loss handling strategies.
\end{enumerate}
Compared to previous approaches, we add support for the emerging Mixture-of-Experts (MoE) architecture and Expert Parallelism (EP), and models long-distance communication time based on real testbed measurements. Our work also examines model chunk sizes within pipeline stages, leverages hierarchical DP communication patterns, and integrates network technologies previously overlooked.
\sysname output include parallelism placement order, job configuration (degrees of parallelisms, microbatch size, model chunk sizes, and communication pattern), pipeline scheduling, and network-level technique recommendations (load balancing and congestion control). 

\subsection{Search Space and Heuristics}

\sysname explores the space of possible combination of parallelism configuration and microbatch size, each defined by a tuple $(TP, CP, EP, PP, DP$ degrees, microbatch size, model chunk sizes). The total number of possible combinations can easily exceed trillions (for example, for a model with 100 layers and 20 pipeline stages, the number of possible layer partitions alone can reach trillions). To make this tractable, we employ several heuristics:

(1) TP is placed at the innermost CP at the second to the innermost layer. TP degree defaulted to the number of accelerators per server, but user-configurable. We bounded CP degree so that each context shard has at least 2,048 tokens.

(2) The difference between the largest and smallest chunk is capped (default 3$\times$) to prevent straggler effects. This bound is user-configurable. 

(3) We perform early pruning for PP-out. Search starts with the minimum number of PP stages required by resource constraints, and stops if further increases in stage count does not improve iteration time.

\begin{figure*}[t]
\begin{minipage}{0.47\textwidth}
\centering
\includegraphics[width=\linewidth]{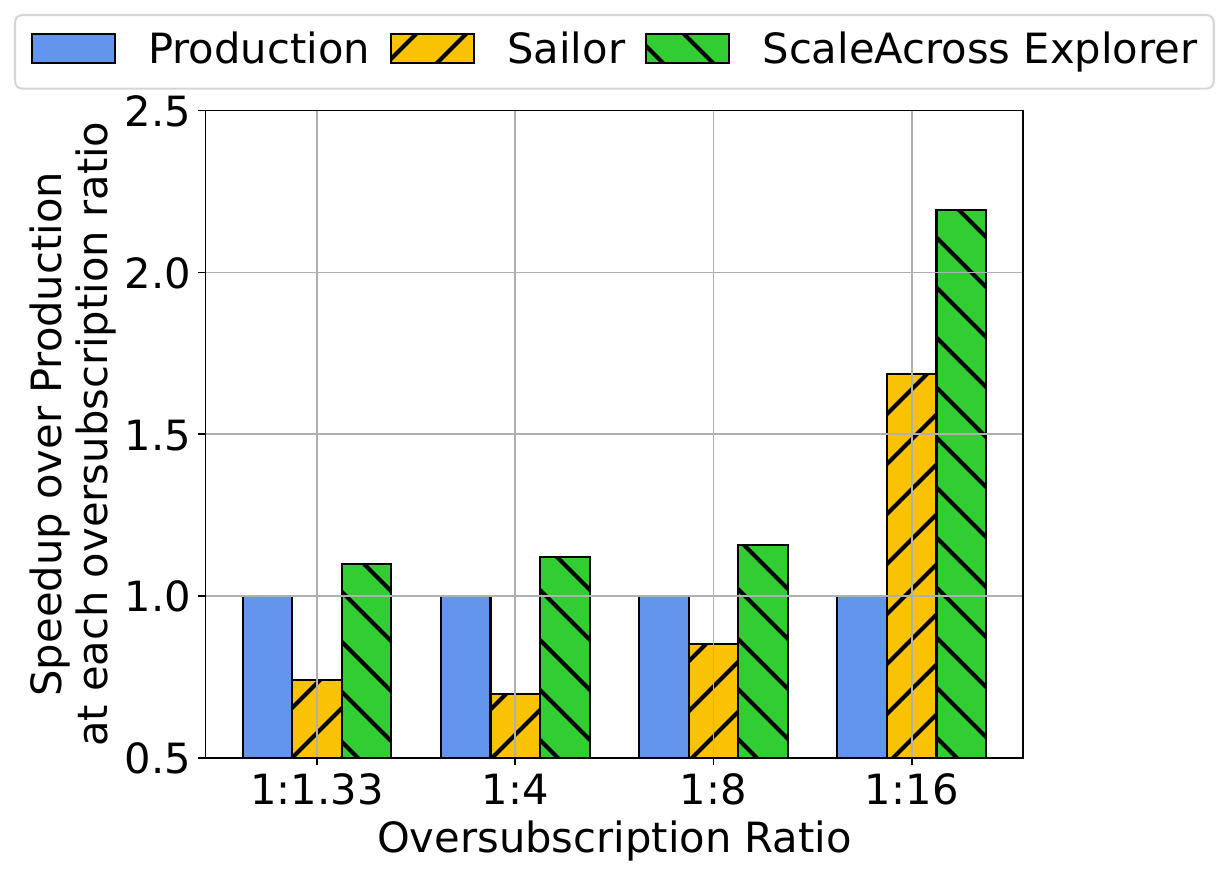}
\captionof{figure}{\small [Testbed] 40B-MoE model training iteration speedup over the production configuration under various oversubscription ratios. 
At a high oversubscription ratio, \sysname achieves a $2.19\times$ speedup over the production baseline while Sailor achieves $1.69\times$.
}
\label{fig:optimize_bdw}
\end{minipage}
\hfill
\begin{minipage}{0.47\textwidth}
    \centering
    \includegraphics[width=\linewidth, trim={0, 0, 0, 0.4cm}, clip]{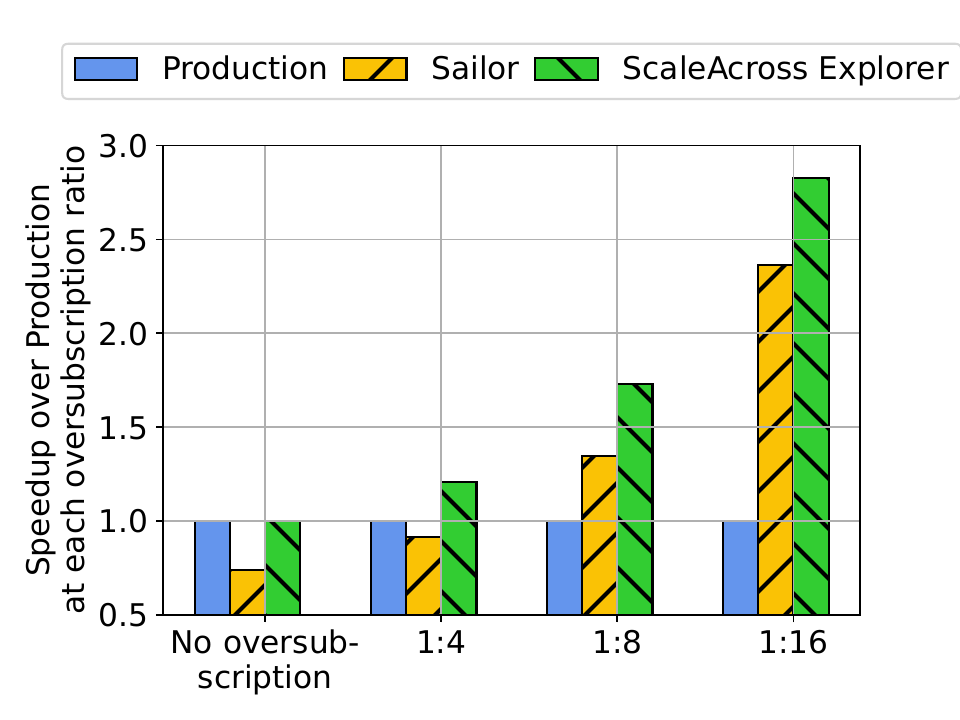}
    \caption{\small [Testbed] 17B-Dense model training iteration speedup over the production configuration under various oversubscription ratios. At a high oversubscription ratio, \sysname achieves a $2.83\times$ speedup over the production baseline while Sailor achieves $2.36\times$.}
    \label{fig:optimize_dense_bdw}
\end{minipage}
\end{figure*}


Despite these heuristics, the main complexity from partitioning layers across pipeline stages and then model chunks remains. The combinatorial nature of the partitioning problem leads to an exponential growth in the search space. To accelerate the search process, we adopt Monte Carlo sampling~\citep{monte_carlo_sampling}, commonly used in recent machine learning system works~\citep{prism,meta_ax,monte_carlo_reasoning}. The sampling process starts with an exploration phase where we randomly sample partitions of layers across stages, evaluate the iteration time of each sampled partition with multiple (100 by default) model chunk configurations and placements on the input topology, and extract features (e.g., largest chunk size, chunk size variation) that correlate with iteration time to guide future sampling. For each of the top $K$ (1000 by default) fastest configurations found through exploration, we start the exploitation phase and generate $M$ (100 by default) new configurations by perturbing model chunks. This approach enables the optimizer to efficiently identify high-quality configurations, accounting for non-linear objectives introduced by pipeline scheduling and chunking, and outperforming traditional random sampling or Mixed-Integer Linear Programming methods.

\begin{figure}[t]
    \centering
    \begin{subfigure}[h]{0.47\textwidth}
        \centering
        \includegraphics[width=\linewidth]{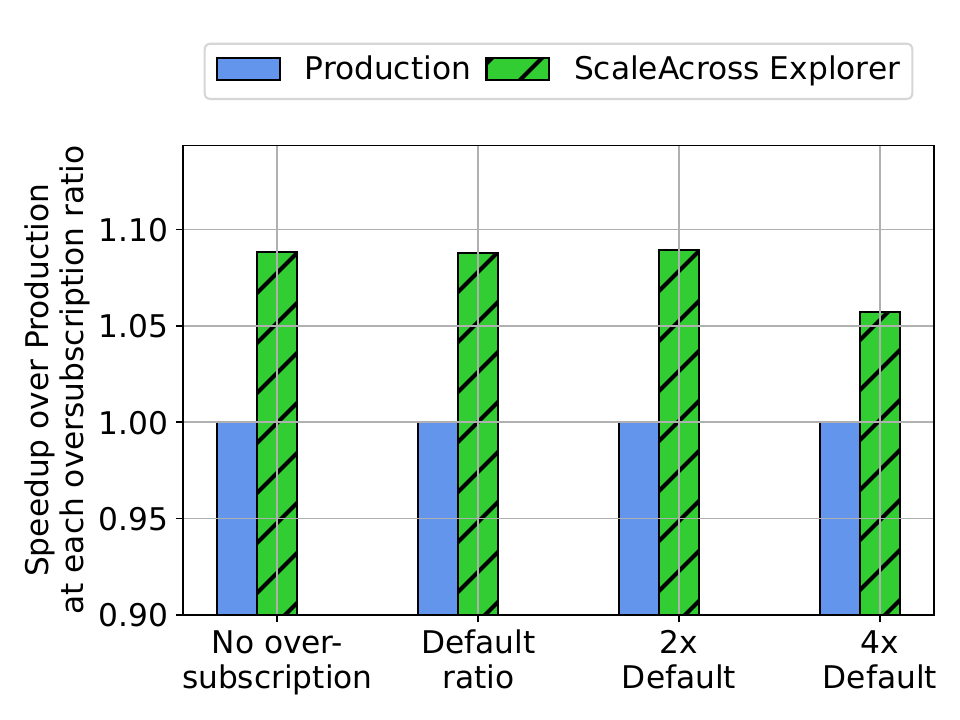}
        \caption{Varying bandwidth oversubscription.}
        \label{fig:optimize_98K_bdw}
    \end{subfigure}
    \begin{subfigure}[h]{0.46\textwidth}
        \centering
        \vspace{0.2cm}
        \includegraphics[width=\linewidth]{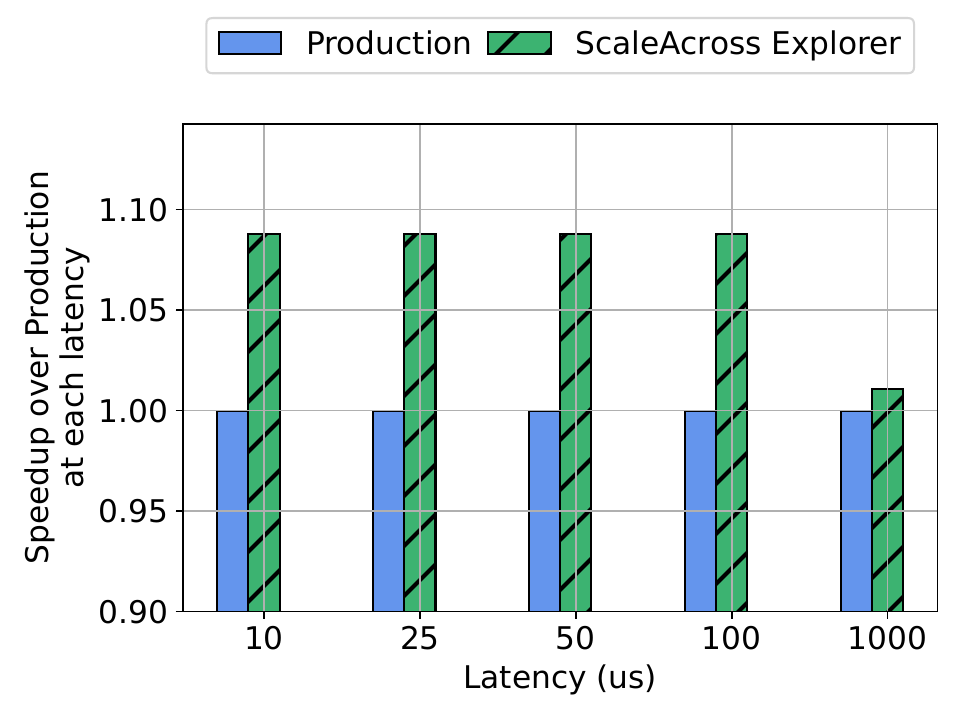}
        \caption{Varying link latencies.}
        \label{fig:optimize_98K_latency}
    \end{subfigure}
    \caption{\small [Simulation] Training iteration speedup over the production trace configuration.  \sysname achieves an up to $1.09\times$ speedup over the production baseline under various oversubscription ratios and latencies. Sailor is not optimized for long-context training and performs poorly; we omit it here to avoid skewing the bars.}
    \label{fig:optimize_98K_nosailor}
\vspace{-0.2cm}
\end{figure}

\begin{figure}[t!]
    \centering
    \begin{subfigure}[h]{0.46\textwidth}
        \centering
        \includegraphics[width=\linewidth]{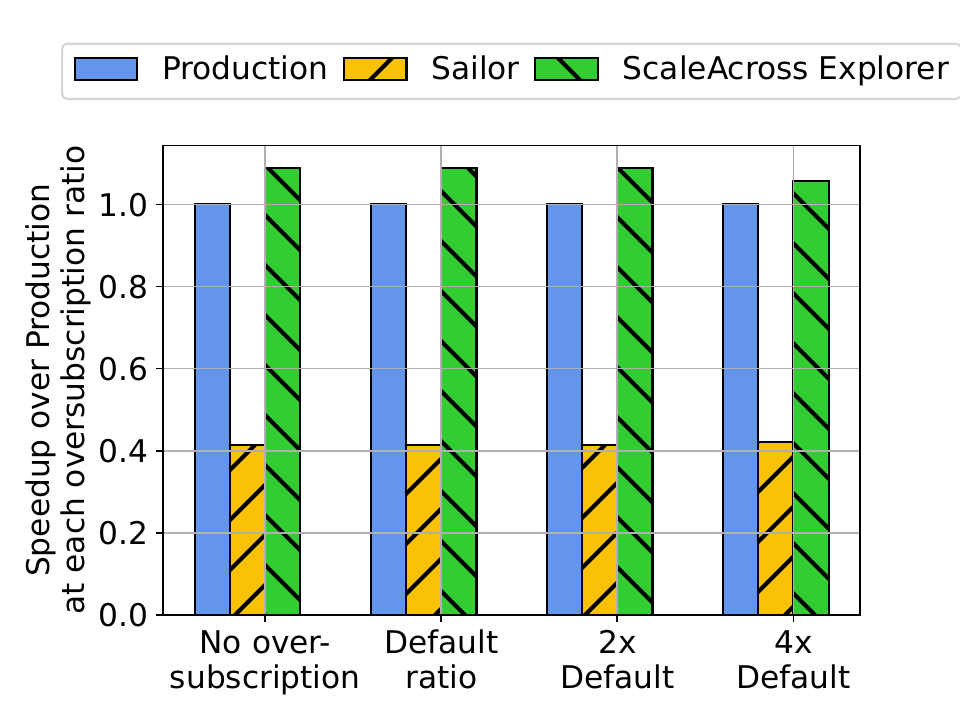}
        \caption{Varying bandwidth oversubscription, with Sailor bars.
        }
        \label{fig:optimize_98K_bdw_withsailor}
    \end{subfigure}
    \begin{subfigure}[h]{0.47\textwidth}
        \centering
        \vspace{0.2cm}
        \includegraphics[width=\linewidth, trim={0, 0, 0, 0.3cm}, clip]{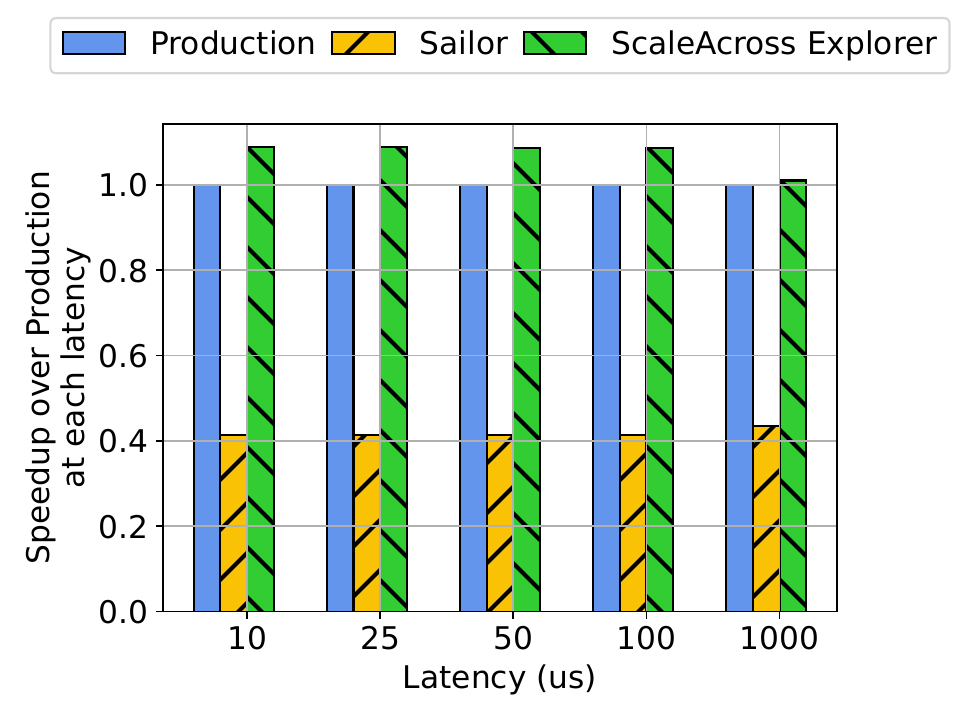}
        \caption{Varying link latencies, with Sailor bars.}
        \label{fig:optimize_98K_latency_withsailor}
    \end{subfigure}
    \caption{\small [Simulation] Training iteration speedup over the production trace configuration, including Sailor results. A speedup less than 1.0 signals that the training iteration becomes longer than that of the production baseline. Sailor is at least 2.32$\times$ slower than the production baseline under various oversubscription ratios and latencies.}
    \label{fig:optimize_98K_withsailor}
\vspace{-0.2cm}
\end{figure}


\subsection{Evaluation Results}
\parab{Testbed.} We first evaluate \sysname on the production testbed.
We use 64 GPUs per cluster and set the cluster distance as 10 kilometers. We compare the iteration time required to train the 40B-MoE model using three configurations: the configuration produced by \sysname, the configuration produced by a baseline training workload optimizer Sailor~\citep{sailor}, and (3) the default production configuration manually tuned by engineers. Batch size is set based on the production configuration.  Figure~\ref{fig:optimize_bdw} demonstrates that \sysname achieves 8.93\% to 54.39\% training iteration time speedup. This improvement is primarily attributed to (1) reducing the number of microbatches per pipeline by optimizing parallelism degrees, (2) switching between DP-out and PP-out as appropriate, and (3) selecting larger model chunk sizes to reduce cross-cluster P2P communication overhead.

\sysname also achieves a 23.07\% to 37.59\% speedup over Sailor. Sailor assumes PP as the outermost parallelism, and optimizes iteration time across valid microbatch sizes and layers per pipeline stage.\footnote{We set throughput as Sailor's optimization objective.} 
However, Sailor only supports the 1F1B (One Forward One Backward) Schedule~\citep{pipedream_1f1b} and overlooks model chunking (i.e., all layers in a pipeline stage are processed together as one model chunk). Therefore, at lower oversubscription ratio (below 1:16) where the performance gap of DP-out and PP-out is narrower, Sailor is slower than the production configuration by 17.40\% to 43.16\% due to pipeline schedule inefficiency.

We also evaluate with training the 17B-Dense model in our research cluster. Due to the small batch size set by the production setting we compared with, PP-out is the better parallelism placement when the network becomes oversubscribed. As shown in Figure~\ref{fig:optimize_dense_bdw}, \sysname achieves a 0.12\% to 64.62\% speedup over the production configuration and a 16.38\% to 26.21\% speedup over Sailor. This improvement comes from identifying PP-out as the better placement and choosing a larger microbatch size and model chunk size to reduce the cross-cluster P2P communication cost. Sailor produces a configuration with eight pipeline stages and without model chunking, leading to poor performance due to the long pipeline warm-up and cool-down periods.

\parab{Simulation.}
For large-scale simulation, we compare our solution with the configuration of the large-scale production run with 100K GPUs.
The large-scale production run is configured with DP at the outermost layer, which is the optimal placement. Nevertheless, as shown in Figure~\ref{fig:optimize_98K_nosailor}, \sysname consistently achieves a speedup of 1.04\% to 8.10\% in iteration time across various oversubscription ratios and latencies. This improvement stems from optimizing intra-building PP by reducing the number of stages, thereby minimizing GPU idle time during pipeline warm-up. The speedups \sysname achieves reduces as latencies increase (1.04\% under 1,000~$\mu$s latency). This is due to \sysname returning a configuration with fewer PP stages and larger DP group size than the production configuration. As discussed in Section~\ref{subsec:moe_eval}, a larger DP group size makes the configuration more sensitive to longer latencies.

As shown in Figure~\ref{fig:optimize_98K_withsailor}, Sailor performs poorly (2.32-2.63$\times$ slower than \sysname) at this scale. The 100K GPU production workload’s long sequence length forces Sailor to use a high TP degree to meet memory constraints, as it lacks support for CP. This causes substantial TP communication overhead. Moreover, Sailor defaults to PP-out, which is suboptimal compared to DP-out for the 100K GPU workload.




\section{Related Work and Lessons Learned}
\label{related_works}

\parab{From Cross-Building to Heterogeneous Scale-Across.}
There are prior work on geo-distributed training. Sailor~\citep{sailor} optimizes the number of and size of pipeline stages and micro batch size within each region for 1F1B scheduling.
CrossPipe~\citep{crosspipe} generates optimized pipeline schedule for a given topology. 
FusionLLM~\citep{fusionllm} supports DP as the outermost parallelism in geo-distributed training and accelerate communication through compression. \sysname considers full stack configurations and supports MoE models. 
This paper primarily characterizes the important communication aspects of training across two data center buildings. Going forward, scale-across training can expand to many data center buildings. These buildings may house different generations and numbers of accelerators and intra-building topologies, and may have heterogeneous distances. Moreover, design decisions extend beyond distributed training efficiency~\cite{madmax} to include availability and reliability~\cite{hpca-2025:reliability}. We therefore need to extend \sysname to such heterogeneous settings.

\parab{Evolvability.}
Having systems that can adapt quickly is critical in the era of large language models, as model architectures and hardware specifications change rapidly. It is prohibitively expensive to run scale-across training for many configurations. Before \sysname, we relied on roofline analysis to determine configurations. However, our experience shows that this approach failed to capture complex yet critical factors, such as, pipeline scheduling and network protocols. \sysname enables agile design space exploration quantitatively and holistically across the model-system-infrastructure stack for distributed training. Having the ability to carry out what-if analysis for the highly-complex design space of distributed training, to characterize the interplay between multiple key design dimensions, and to determine higher-performing training workload configurations can significantly reduce experimentation and LLM development cost.  

A key enabler for \sysname is our configurable scale-across testbed featuring optical systems. This testbed was essential not just to inform and validate the evaluation results, but it also helped uncover non-intuitive network behavior in NIC and switch implementations. 

\parab{Interplay between System Efficiency and Training Algorithms.}
A key design choice in \sysname is to push the limits of infrastructure efficiency without affecting model accuracy. We deliberately focus on lossless optimizations—such as parallelism placement and scheduling (e.g., DoraPP, Interleaved ZBV)—that result in mathematically equivalent training to a single-cluster baseline.

We note that there are many research on asynchronous training. Local SGD leverages nearby GPUs from local groups to train a local model synchronously, and periodically communicate their local model with other groups~\citep{async_localsgd, diloco, localsgd}. HALoS~\citep{halos} further accelerates Local SGD by removing the synchronization barrier of local models. StellaTrain~\citep{stellatrain} studies semi-asynchronous training with partially stale layer parameter updates.
\sysname will consider these alternative training algorithm options once they become production ready. 

\textbf{Communication Collective and Network Optimization.}
Advanced collective schedules, such as Parallel Aggregated Trees (PAT~\citep{nccl_pat}) and Bruck algorithms~\citep{bruckalgo}, 
offer logarithmic latency benefits but introduce significant implementation complexity due to intricate buffer management and non-contiguous memory access patterns. Furthermore, the computational overhead required for local data packing and unpacking in these algorithms can potentially negate network latency gains if not highly optimized for specific hardware architectures. This is an important area of research. We need to co-design next-generation networking protocols and collective algorithms for fast, efficient cross-building communications to enable more network-layer design choices for \sysname.

\section{Conclusion}
In this work, we present a comprehensive exploration of the communication optimization strategies for training workloads that scale across multiple buildings, data centers, and geographic regions. Our study reveals that the optimal parallelism placement depends on model architecture and job configuration, while careful model chunk sizing and hierarchical communication patterns can substantially accelerate training. We show that pipeline schedules should be tuned to minimize communication frequency over the constrained scaling-across network. We also demonstrate the impact of long link latency on lossy network and the effectiveness of load balancing and congestion control. Building on these insights, we developed a workload optimizer that consistently outperforms existing solutions.

\clearpage
\newpage
\bibliographystyle{assets/plainnat}

\begin{thebibliography}{56}
\providecommand{\natexlab}[1]{#1}
\providecommand{\url}[1]{\texttt{#1}}
\expandafter\ifx\csname urlstyle\endcsname\relax
  \providecommand{\doi}[1]{doi: #1}\else
  \providecommand{\doi}{doi: \begingroup \urlstyle{rm}\Url}\fi

\bibitem[Bagga et~al.(2025)Bagga, Fang, Sunkad, Puri, Moeller, Wu, Vijayanath, Vasudevan, and Jani]{ocp_summit_2025}
Jasmeet Bagga, Tian Fang, Ravindra Sunkad, Rohit Puri, Olaf Moeller, Lingjun Wu, Vignesh Vijayanath, Vimal Vasudevan, and Dharmesh Jani.
\newblock Ocp summit 2025: The open future of networking hardware for ai, 2025.
\newblock \url{https://engineering.fb.com/2025/10/13/data-infrastructure/ocp-summit-2025-the-open-future-of-networking-hardware-for-ai/}.
\newblock Accessed: 2026-02-02.

\bibitem[Bruck et~al.(1994)Bruck, Ho, Kipnis, and Weathersby]{bruckalgo}
Jehoshua Bruck, Ching-Tien Ho, Shlomo Kipnis, and Derrick Weathersby.
\newblock Efficient algorithms for all-to-all communications in multi-port message-passing systems.
\newblock In \emph{Proceedings of the sixth annual ACM symposium on Parallel algorithms and architectures}, pages 298--309, 1994.

\bibitem[Chen et~al.(2025)Chen, Kubicek, Huang, and Hoefler]{crosspipe}
Tiancheng Chen, Ales Kubicek, Langwen Huang, and Torsten Hoefler.
\newblock Crosspipe: towards optimal pipeline schedules for cross-datacenter training.
\newblock In \emph{Proceedings of the 2025 USENIX Conference on Usenix Annual Technical Conference}, USENIX ATC '25, USA, 2025. USENIX Association.
\newblock ISBN 978-1-939133-48-9.

\bibitem[Choukse et~al.(2025)Choukse, Warrier, Heath, Belmont, Zhao, Khan, Harry, Kappel, Hewett, Datta, et~al.]{ai_power_stabilization}
Esha Choukse, Brijesh Warrier, Scot Heath, Luz Belmont, April Zhao, Hassan~Ali Khan, Brian Harry, Matthew Kappel, Russell~J Hewett, Kushal Datta, et~al.
\newblock Power stabilization for ai training datacenters.
\newblock \emph{arXiv preprint arXiv:2508.14318}, 2025.

\bibitem[Chu et~al.(2025)Chu, Xie, Yu, Wang, Phanishayee, Tang, Hao, Huang, Ozdal, Wang, et~al.]{scaling_llama3}
Weiwei Chu, Xinfeng Xie, Jiecao Yu, Jie Wang, Amar Phanishayee, Chunqiang Tang, Yuchen Hao, Jianyu Huang, Mustafa Ozdal, Jun Wang, et~al.
\newblock Scaling llama 3 training with efficient parallelism strategies.
\newblock In \emph{Proceedings of the 52nd Annual International Symposium on Computer Architecture}, pages 1703--1716, 2025.

\bibitem[Corporation(2023)]{nvidia_connectx6dx}
NVIDIA Corporation.
\newblock Nvidia connectx-6 dx adapter cards firmware release notes v22.35.3006 lts, 2023.
\newblock \url{https://docs.nvidia.com/networking/display/connectx6dxfirmwarev22353006lts/changes+and+new+feature+history}.
\newblock Accessed: 2025-12-18.

\bibitem[Dean and Barroso(2013)]{tail_at_scale}
Jeffrey Dean and Luiz~Andr{\'e} Barroso.
\newblock The tail at scale.
\newblock \emph{Communications of the ACM}, 56\penalty0 (2):\penalty0 74--80, 2013.

\bibitem[Dixit et~al.(2013{\natexlab{a}})Dixit, Prakash, Hu, and Kompella]{impact_packet_spray}
Advait Dixit, Pawan Prakash, Y~Charlie Hu, and Ramana~Rao Kompella.
\newblock On the impact of packet spraying in data center networks.
\newblock In \emph{2013 proceedings ieee infocom}, pages 2130--2138. IEEE, 2013{\natexlab{a}}.

\bibitem[Dixit et~al.(2013{\natexlab{b}})Dixit, Prakash, Hu, and Kompella]{packet_spray}
Advait Dixit, Pawan Prakash, Y.~Charlie Hu, and Ramana~Rao Kompella.
\newblock On the impact of packet spraying in data center networks.
\newblock In \emph{2013 Proceedings IEEE INFOCOM}, pages 2130--2138, 2013{\natexlab{b}}.
\newblock \doi{10.1109/INFCOM.2013.6567015}.

\bibitem[Douillard et~al.(2024)Douillard, Feng, Rusu, Chhaparia, Donchev, Kuncoro, Ranzato, Szlam, and Shen]{diloco}
Arthur Douillard, Qixuan Feng, Andrei~A. Rusu, Rachita Chhaparia, Yani Donchev, Adhiguna Kuncoro, Marc'Aurelio Ranzato, Arthur Szlam, and Jiajun Shen.
\newblock Diloco: Distributed low-communication training of language models.
\newblock 2024.
\newblock \url{https://arxiv.org/abs/2311.08105}.

\bibitem[Floyd et~al.(2001)Floyd, Ramakrishnan, and Black]{ecn}
Sally Floyd, Dr. K.~K. Ramakrishnan, and David~L. Black.
\newblock {The Addition of Explicit Congestion Notification (ECN) to IP}.
\newblock RFC 3168, September 2001.
\newblock \url{https://www.rfc-editor.org/info/rfc3168}.

\bibitem[Gandhi et~al.(2024)Gandhi, Tandon, Bhattacherjee, Padmanabhan, et~al.]{microsoft_geo}
Rohan Gandhi, Karan Tandon, Debopam Bhattacherjee, Venkata~N Padmanabhan, et~al.
\newblock Improving training time and gpu utilization in geo-distributed language model training.
\newblock \emph{arXiv preprint arXiv:2411.14458}, 2024.

\bibitem[Gangidi et~al.(2024)Gangidi, Miao, Zheng, Bondu, Goes, Morsy, Puri, Riftadi, Shetty, Yang, et~al.]{meta_roce_at_scale}
Adithya Gangidi, Rui Miao, Shengbao Zheng, Sai~Jayesh Bondu, Guilherme Goes, Hany Morsy, Rohit Puri, Mohammad Riftadi, Ashmitha~Jeevaraj Shetty, Jingyi Yang, et~al.
\newblock Rdma over ethernet for distributed training at meta scale.
\newblock In \emph{Proceedings of the ACM SIGCOMM 2024 Conference}, pages 57--70, 2024.

\bibitem[Gherghescu et~al.(2024)Gherghescu, B{\u{a}}doiu, Agache, Dumitru, Vasilescu, Mantu, and Raiciu]{ml_dc_network_research}
Alexandru~M Gherghescu, Vlad-Andrei B{\u{a}}doiu, Alexandru Agache, Mihai-Valentin Dumitru, Iuliu Vasilescu, Radu Mantu, and Costin Raiciu.
\newblock I've got 99 problems but flops ain't one.
\newblock In \emph{Proceedings of the 23rd ACM Workshop on Hot Topics in Networks}, pages 195--204, 2024.

\bibitem[Golden et~al.(2025)Golden, Kuchnik, Hsia, DeVito, Wei, Brooks, and Wu]{prism}
Alicia Golden, Michael Kuchnik, Samuel Hsia, Zachary DeVito, Gu-Yeon Wei, David Brooks, and Carole-Jean Wu.
\newblock Prism: Probabilistic runtime insights and scalable performance modeling for large-scale distributed training.
\newblock \emph{arXiv preprint arXiv:2510.15596}, 2025.

\bibitem[Grattafiori et~al.(2024)Grattafiori, Dubey, Jauhri, Pandey, Kadian, Al-Dahle, Letman, Mathur, Schelten, Vaughan, et~al.]{llama3}
Aaron Grattafiori, Abhimanyu Dubey, Abhinav Jauhri, Abhinav Pandey, Abhishek Kadian, Ahmad Al-Dahle, Aiesha Letman, Akhil Mathur, Alan Schelten, Alex Vaughan, et~al.
\newblock The llama 3 herd of models.
\newblock \emph{arXiv preprint arXiv:2407.21783}, 2024.

\bibitem[Guo et~al.(2016)Guo, Wu, Deng, Soni, Ye, Padhye, and Lipshteyn]{rdma_gobackn}
Chuanxiong Guo, Haitao Wu, Zhong Deng, Gaurav Soni, Jianxi Ye, Jitu Padhye, and Marina Lipshteyn.
\newblock Rdma over commodity ethernet at scale.
\newblock In \emph{Proceedings of the 2016 ACM SIGCOMM Conference}, pages 202--215, 2016.

\bibitem[Guo et~al.(2025)Guo, Wu, Zhu, Leng, Shi, Chen, Fan, Wang, Jiang, Wang, et~al.]{bytedance_seed}
Dong Guo, Faming Wu, Feida Zhu, Fuxing Leng, Guang Shi, Haobin Chen, Haoqi Fan, Jian Wang, Jianyu Jiang, Jiawei Wang, et~al.
\newblock Seed1.5-vl technical report.
\newblock \emph{arXiv preprint arXiv:2505.07062}, 2025.

\bibitem[Hopps(2000)]{ecmp}
Christian Hopps.
\newblock Analysis of an equal-cost multi-path algorithm.
\newblock Technical Report RFC 2992, IETF, 2000.
\newblock \url{https://datatracker.ietf.org/doc/html/rfc2992}.

\bibitem[Hsia et~al.(2024)Hsia, Golden, Acun, Ardalani, DeVito, Wei, Brooks, and Wu]{madmax}
Samuel Hsia, Alicia Golden, Bilge Acun, Newsha Ardalani, Zachary DeVito, Gu-Yeon Wei, David Brooks, and Carole-Jean Wu.
\newblock Mad-max beyond single-node: Enabling large machine learning model acceleration on distributed systems.
\newblock In \emph{2024 ACM/IEEE 51st Annual International Symposium on Computer Architecture (ISCA)}, pages 818--833. IEEE, 2024.

\bibitem[Jeaugey(2025)]{nccl_pat}
Sylvain Jeaugey.
\newblock Pat: a new algorithm for all-gather and reduce-scatter operations at scale, 2025.
\newblock \url{https://arxiv.org/abs/2506.20252}.

\bibitem[Khalilov et~al.(2025)Khalilov, Shen, Chrapek, Chen, Nakano, Mazzoletti, Gootzen, Di~Girolamo, Nudelman, Bloch, et~al.]{msft_sdr}
Mikhail Khalilov, Siyuan Shen, Marcin Chrapek, Tiancheng Chen, Kenji Nakano, Nicola Mazzoletti, Peter-Jan Gootzen, Salvatore Di~Girolamo, Rami Nudelman, Gil Bloch, et~al.
\newblock Sdr-rdma: Software-defined reliability architecture for planetary scale rdma communication.
\newblock In \emph{Proceedings of the International Conference for High Performance Computing, Networking, Storage and Analysis}, pages 1223--1239, 2025.

\bibitem[Kim et~al.(2025)Kim, Li, Gandham, Baldonado, Gangidi, Balaji, Wang, and Akella]{halos}
Geon-Woo Kim, Junbo Li, Shashidhar Gandham, Omar Baldonado, Adithya Gangidi, Pavan Balaji, Zhangyang Wang, and Aditya Akella.
\newblock {HAL}os: Hierarchical asynchronous local {SGD} over slow networks for geo-distributed large language model training.
\newblock In \emph{Forty-second International Conference on Machine Learning}, 2025.
\newblock \url{https://openreview.net/forum?id=LMPJnZSNC8}.

\bibitem[Kokolis et~al.(2025)Kokolis, Kuchnik, Hoffman, Kumar, Malani, Ma, DeVito, Sengupta, Saladi, and Wu]{hpca-2025:reliability}
Apostolos Kokolis, Michael Kuchnik, John Hoffman, Adithya Kumar, Parth Malani, Faye Ma, Zachary DeVito, Shubho Sengupta, Kalyan Saladi, and Carole-Jean Wu.
\newblock { Revisiting Reliability in Large-Scale Machine Learning Research Clusters }.
\newblock In \emph{2025 IEEE International Symposium on High Performance Computer Architecture (HPCA)}, 2025.

\bibitem[Lim et~al.(2024)Lim, Ye, Abdu~Jyothi, and Han]{stellatrain}
Hwijoon Lim, Juncheol Ye, Sangeetha Abdu~Jyothi, and Dongsu Han.
\newblock Accelerating model training in multi-cluster environments with consumer-grade gpus.
\newblock In \emph{Proceedings of the ACM SIGCOMM 2024 Conference}, ACM SIGCOMM '24, page 707–720, New York, NY, USA, 2024. Association for Computing Machinery.
\newblock ISBN 9798400706141.
\newblock \doi{10.1145/3651890.3672228}.
\newblock \url{https://doi.org/10.1145/3651890.3672228}.

\bibitem[Lin et~al.(2020)Lin, Stich, Patel, and Jaggi]{localsgd}
Tao Lin, Sebastian~U. Stich, Kumar~Kshitij Patel, and Martin Jaggi.
\newblock Don't use large mini-batches, use local sgd.
\newblock In \emph{International Conference on Learning Representations}, 2020.
\newblock \url{https://openreview.net/forum?id=B1eyO1BFPr}.

\bibitem[Liu et~al.(2024{\natexlab{a}})Liu, Feng, Xue, Wang, Wu, Lu, Zhao, Deng, Zhang, Ruan, et~al.]{deepseekv3}
Aixin Liu, Bei Feng, Bing Xue, Bingxuan Wang, Bochao Wu, Chengda Lu, Chenggang Zhao, Chengqi Deng, Chenyu Zhang, Chong Ruan, et~al.
\newblock Deepseek-v3 technical report.
\newblock \emph{arXiv preprint arXiv:2412.19437}, 2024{\natexlab{a}}.

\bibitem[Liu et~al.(2024{\natexlab{b}})Liu, Chhaparia, Douillard, Kale, Rusu, Shen, Szlam, and Ranzato]{async_localsgd}
Bo~Liu, Rachita Chhaparia, Arthur Douillard, Satyen Kale, Andrei~Alex Rusu, Jiajun Shen, Arthur Szlam, and MarcAurelio Ranzato.
\newblock Asynchronous local-{SGD} training for language modeling.
\newblock In \emph{2nd Workshop on Advancing Neural Network Training: Computational Efficiency, Scalability, and Resource Optimization (WANT@ICML 2024)}, 2024{\natexlab{b}}.
\newblock \url{https://openreview.net/forum?id=mT3PdRKl40}.

\bibitem[Liu et~al.(2025{\natexlab{a}})Liu, Su, Yao, Jiang, Lai, Du, Qin, Xu, Lu, Yan, et~al.]{moonshot_moonlight}
Jingyuan Liu, Jianlin Su, Xingcheng Yao, Zhejun Jiang, Guokun Lai, Yulun Du, Yidao Qin, Weixin Xu, Enzhe Lu, Junjie Yan, et~al.
\newblock Muon is scalable for llm training.
\newblock \emph{arXiv preprint arXiv:2502.16982}, 2025{\natexlab{a}}.

\bibitem[Liu et~al.(2025{\natexlab{b}})Liu, Li, and Chen]{enabling_packet_spray}
Xiangzhou Liu, Wenxue Li, and Kai Chen.
\newblock Enabling packet spraying over commodity rnics with in-network support.
\newblock In \emph{Proceedings of the 9th Asia-Pacific Workshop on Networking}, pages 51--58, 2025{\natexlab{b}}.

\bibitem[M2~Optics(2026)]{m2optics_fiberlab}
Inc. M2~Optics.
\newblock Fiber network and link simulation solutions.
\newblock \url{https://www.m2optics.com/solutions/network-simulation-latency}, 2026.
\newblock Accessed: 2026-02-05.

\bibitem[Meta(2025)]{atscale2025metadc}
Meta.
\newblock Meta’s dc networks for generative ai, 2025.
\newblock \url{https://atscaleconference.com/videos/metas-dc-networks-for-generative-ai/}.
\newblock AtScale Conference. Accessed: 2025-12-09.

\bibitem[Metropolis et~al.(1953)Metropolis, Rosenbluth, Rosenbluth, Teller, and Teller]{monte_carlo_sampling}
Nicholas Metropolis, Arianna~W Rosenbluth, Marshall~N Rosenbluth, Augusta~H Teller, and Edward Teller.
\newblock Equation of state calculations by fast computing machines.
\newblock \emph{The journal of chemical physics}, 21\penalty0 (6):\penalty0 1087--1092, 1953.

\bibitem[Morales(2024)]{tomshardware2024chinaai}
Jowi Morales.
\newblock China makes ai breakthrough, reportedly trains generative ai model across multiple data centers and gpu architectures, 2024.
\newblock \url{https://www.tomshardware.com/tech-industry/artificial-intelligence/china-makes-ai-breakthrough-reportedly-trains-generative-ai-model-across-multiple-data-centers-and-gpu-architectures}.
\newblock Tom's Hardware. Accessed: 2025-12-09.

\bibitem[Narayanan et~al.(2019)Narayanan, Harlap, Phanishayee, Seshadri, Devanur, Ganger, Gibbons, and Zaharia]{pipedream_1f1b}
Deepak Narayanan, Aaron Harlap, Amar Phanishayee, Vivek Seshadri, Nikhil~R. Devanur, Gregory~R. Ganger, Phillip~B. Gibbons, and Matei Zaharia.
\newblock Pipedream: generalized pipeline parallelism for dnn training.
\newblock In \emph{Proceedings of the 27th ACM Symposium on Operating Systems Principles}, SOSP '19, page 1–15, New York, NY, USA, 2019. Association for Computing Machinery.
\newblock ISBN 9781450368735.
\newblock \doi{10.1145/3341301.3359646}.
\newblock \url{https://doi.org/10.1145/3341301.3359646}.

\bibitem[Olson et~al.(2025)Olson, Santorella, Tiao, Cakmak, Garrard, Daulton, Lin, Ament, Beckerman, Onofrey, et~al.]{meta_ax}
Miles Olson, Elizabeth Santorella, Louis~C Tiao, Sait Cakmak, Mia Garrard, Samuel Daulton, Zhiyuan~Jerry Lin, Sebastian Ament, Bernard Beckerman, Eric Onofrey, et~al.
\newblock Ax: a platform for adaptive experimentation.
\newblock In \emph{AutoML 2025 ABCD Track}, 2025.

\bibitem[Ontiveros et~al.(2023)Ontiveros, Patel, and Pandey]{semianalysis_grid_power_blackout}
Jeremie~Eliahou Ontiveros, Dylan Patel, and Ajey Pandey.
\newblock Ai training load fluctuations at gigawatt scale: Risk of power grid blackout, 2023.
\newblock \url{https://newsletter.semianalysis.com/p/ai-training-load-fluctuations-at-gigawatt-scale-risk-of-power-grid-blackout}.
\newblock Accessed: 2025-12-04.

\bibitem[Patel et~al.(2025)Patel, Nishball, and Ontiveros]{semianalysis_multidc}
Dylan Patel, Daniel Nishball, and Jeremie~Eliahou Ontiveros.
\newblock Multi datacenter training: Openai’s secret sauce for scaling ai, 2025.
\newblock \url{https://newsletter.semianalysis.com/p/multi-datacenter-training-openais}.
\newblock Accessed: 2025-12-05.

\bibitem[Qi et~al.(2024)Qi, Wan, Huang, and Lin]{zerobubble}
Penghui Qi, Xinyi Wan, Guangxing Huang, and Min Lin.
\newblock Zero bubble (almost) pipeline parallelism.
\newblock In \emph{The Twelfth International Conference on Learning Representations}, 2024.
\newblock \url{https://openreview.net/forum?id=tuzTN0eIO5}.

\bibitem[Rajbhandari et~al.(2021)Rajbhandari, Ruwase, Rasley, Smith, and He]{zero_infinity}
Samyam Rajbhandari, Olatunji Ruwase, Jeff Rasley, Shaden Smith, and Yuxiong He.
\newblock Zero-infinity: Breaking the gpu memory wall for extreme scale deep learning.
\newblock In \emph{Proceedings of the international conference for high performance computing, networking, storage and analysis}, pages 1--14, 2021.

\bibitem[Sadok et~al.(2018)Sadok, Campista, and Costa]{case_for_packet_spray}
Hugo Sadok, Miguel Elias~M Campista, and Lu{\'\i}s Henrique~MK Costa.
\newblock A case for spraying packets in software middleboxes.
\newblock In \emph{Proceedings of the 17th ACM Workshop on Hot Topics in Networks}, pages 127--133, 2018.

\bibitem["@Scale"(2025{\natexlab{a}})]{meta_100k_optimization}
"@Scale".
\newblock Performance optimizations at 100k+ scale by ashmitha jeevaraj shetty and min si, 2025{\natexlab{a}}.
\newblock \url{https://www.youtube.com/watch?v=XoTok_8lFXE}.
\newblock Accessed: 2025-12-15.

\bibitem["@Scale"(2025{\natexlab{b}})]{meta_dc_networks}
"@Scale".
\newblock Meta’s dc networks for generative ai by rohit puri and hany morsy, 2025{\natexlab{b}}.
\newblock \url{https://www.youtube.com/watch?v=AqIPRseYcTU}.
\newblock Accessed: 2025-12-15.

\bibitem[Shuai et~al.(2024)Shuai, Wang, Wu, Jiang, and Ren]{batchsize_scaling}
Xian Shuai, Yiding Wang, Yimeng Wu, Xin Jiang, and Xiaozhe Ren.
\newblock Scaling law for language models training considering batch size.
\newblock \emph{arXiv preprint arXiv:2412.01505}, 2024.

\bibitem[Si et~al.(2026)Si, Balaji, Chen, Chu, Gangidi, Hasan, Iyengar, Johnson, Liu, Ren, Shah, Shetty, Steinbrecher, Wang, Wu, Xie, Yang, Yang, Yu, Yu, Zhao, Bland, Boyda, Gumudavelli, Kannan, Lumezanu, Miao, Qu, Ramesh, Samoylov, Seidel, Sundaresan, Tian, Tan, Zhang, Zhao, Zheng, Zhu, and Zeng]{ncclx_arxiv}
Min Si, Pavan Balaji, Yongzhou Chen, Ching-Hsiang Chu, Adi Gangidi, Saif Hasan, Subodh Iyengar, Dan Johnson, Bingzhe Liu, Regina Ren, Deep Shah, Ashmitha~Jeevaraj Shetty, Greg Steinbrecher, Yulun Wang, Bruce Wu, Xinfeng Xie, Jingyi Yang, Mingran Yang, Kenny Yu, Minlan Yu, Cen Zhao, Wes Bland, Denis Boyda, Suman Gumudavelli, Prashanth Kannan, Cristian Lumezanu, Rui Miao, Zhe Qu, Venkat Ramesh, Maxim Samoylov, Jan Seidel, Srikanth Sundaresan, Feng Tian, Qiye Tan, Shuqiang Zhang, Yimeng Zhao, Shengbao Zheng, Art Zhu, and Hongyi Zeng.
\newblock Collective communication for 100k+ gpus, 2026.
\newblock \url{https://arxiv.org/abs/2510.20171}.

\bibitem[Song and Veeraraghavan(2025)]{meta2025infrastructure}
Yee~Jiun Song and Kaushik Veeraraghavan.
\newblock Meta’s infrastructure evolution and the advent of ai, 2025.
\newblock \url{https://engineering.fb.com/2025/09/29/data-infrastructure/metas-infrastructure-evolution-and-the-advent-of-ai/}.
\newblock Meta Engineering Blog. Accessed: 2025-12-07.

\bibitem[Sriraman et~al.(2017)Sriraman, Liu, Gunbay, Su, and Wenisch]{deconstructing_tail_at_scale}
Akshitha Sriraman, Sihang Liu, Sinan Gunbay, Shan Su, and Thomas~F Wenisch.
\newblock Deconstructing the tail at scale effect across network protocols.
\newblock \emph{arXiv preprint arXiv:1701.03100}, 2017.

\bibitem[Strati et~al.(2024)Strati, Elvinger, Kerimoglu, and Klimovic]{euromlsys_cross_region}
Foteini Strati, Paul Elvinger, Tolga Kerimoglu, and Ana Klimovic.
\newblock Ml training with cloud gpu shortages: Is cross-region the answer?
\newblock In \emph{Proceedings of the 4th Workshop on Machine Learning and Systems}, pages 107--116, 2024.

\bibitem[Strati et~al.(2025)Strati, Zhang, Manos, P{\'e}riz, Hu, Chen, Buzcu, Han, Delgado, and Klimovic]{sailor}
Foteini Strati, Zhendong Zhang, George Manos, Ixeia~S{\'a}nchez P{\'e}riz, Qinghao Hu, Tiancheng Chen, Berk Buzcu, Song Han, Pamela Delgado, and Ana Klimovic.
\newblock Sailor: Automating distributed training over dynamic, heterogeneous, and geo-distributed clusters.
\newblock In \emph{Proceedings of the ACM SIGOPS 31st Symposium on Operating Systems Principles}, pages 204--220, 2025.

\bibitem[Tang et~al.(2024)Tang, Kang, Yin, Pan, Wang, He, Wang, Zeng, Zhao, Shi, et~al.]{fusionllm}
Zhenheng Tang, Xueze Kang, Yiming Yin, Xinglin Pan, Yuxin Wang, Xin He, Qiang Wang, Rongfei Zeng, Kaiyong Zhao, Shaohuai Shi, et~al.
\newblock Fusionllm: a decentralized llm training system on geo-distributed gpus with adaptive compression.
\newblock \emph{arXiv preprint arXiv:2410.12707}, 2024.

\bibitem[Wang et~al.(2026)Wang, Ahuja, Zhang, Zhang, Noormohammadpour, Steinbrecher, Fuller, Liu, Quirk, Fernandez, Triguna, Cai, Politis, Lapukhov, Hasani, and Zhang]{arcadia}
Zhaodong Wang, Satyajeet~Singh Ahuja, Xu~Zhang, Yuhui Zhang, Max Noormohammadpour, Gregory~R. Steinbrecher, Thomas Fuller, Xin Liu, Kevin Quirk, Mikel~Jimenez Fernandez, Abhinav Triguna, Yan Cai, Steve Politis, Petr Lapukhov, Naader Hasani, and Ying Zhang.
\newblock Enabling ai network cross-layer design and operations with arcadia: A simulation platform at scale.
\newblock In \emph{Accepted for 23rd USENIX Symposium on Networked Systems Design and Implementation (NSDI 26)}, Philadelphia, PA, 2026. USENIX Association.

\bibitem[Xie et~al.(2024)Xie, Goyal, Zheng, Kan, Lillicrap, Kawaguchi, and Shieh]{monte_carlo_reasoning}
Yuxi Xie, Anirudh Goyal, Wenyue Zheng, Min-Yen Kan, Timothy~P Lillicrap, Kenji Kawaguchi, and Michael Shieh.
\newblock Monte carlo tree search boosts reasoning via iterative preference learning.
\newblock \emph{arXiv preprint arXiv:2405.00451}, 2024.

\bibitem[Yang et~al.(2025)Yang, Li, Zeng, and Gu]{fec_for_rdma}
Zhiyi Yang, Zhexiong Li, Deze Zeng, and Lin Gu.
\newblock Dcts-rdma: Adaptive fec via dynamic coding for efficient rdma over lossy networks.
\newblock In \emph{IFIP International Conference on Network and Parallel Computing}, pages 235--246. Springer, 2025.

\bibitem[Zhao et~al.(2023)Zhao, Gu, Varma, Luo, Huang, Xu, Wright, Shojanazeri, Ott, Shleifer, Desmaison, Balioglu, Damania, Nguyen, Chauhan, Hao, Mathews, and Li]{fsdp}
Yanli Zhao, Andrew Gu, Rohan Varma, Liang Luo, Chien-Chin Huang, Min Xu, Less Wright, Hamid Shojanazeri, Myle Ott, Sam Shleifer, Alban Desmaison, Can Balioglu, Pritam Damania, Bernard Nguyen, Geeta Chauhan, Yuchen Hao, Ajit Mathews, and Shen Li.
\newblock Pytorch fsdp: Experiences on scaling fully sharded data parallel.
\newblock \emph{Proc. VLDB Endow.}, 16\penalty0 (12):\penalty0 3848–3860, August 2023.
\newblock ISSN 2150-8097.
\newblock \doi{10.14778/3611540.3611569}.
\newblock \url{https://doi.org/10.14778/3611540.3611569}.

\bibitem[Zhu et~al.(2015)Zhu, Eran, Firestone, Guo, Lipshteyn, Liron, Padhye, Raindel, Yahia, and Zhang]{rdma_congestion}
Yibo Zhu, Haggai Eran, Daniel Firestone, Chuanxiong Guo, Marina Lipshteyn, Yehonatan Liron, Jitendra Padhye, Shachar Raindel, Mohamad~Haj Yahia, and Ming Zhang.
\newblock Congestion control for large-scale rdma deployments.
\newblock \emph{ACM SIGCOMM Computer Communication Review}, 45\penalty0 (4):\penalty0 523--536, 2015.

\bibitem[Zu et~al.(2024)Zu, Ghaffarkhah, Dang, Towles, Hand, Huda, Bello, Kolbasov, Rezaei, Du, et~al.]{google_resiliency}
Yazhou Zu, Alireza Ghaffarkhah, Hoang-Vu Dang, Brian Towles, Steven Hand, Safeen Huda, Adekunle Bello, Alexander Kolbasov, Arash Rezaei, Dayou Du, et~al.
\newblock Resiliency at scale: Managing $\{$Google’s$\}$$\{$TPUv4$\}$ machine learning supercomputer.
\newblock In \emph{21st USENIX Symposium on Networked Systems Design and Implementation (NSDI 24)}, pages 761--774, 2024.

\end{thebibliography}


\clearpage
\newpage
\beginappendix

\section{Exploration Details}

\subsection{Testbed Settings}
\label{app:method_emulator}

We evaluate scale-across training with a production testbed equipped with tools capable of increasing bandwidth oversubscription ratios by disabling links and reconfiguring cluster distances through optical systems. Because the testbed is shared among multiple users, we prioritize experiments with Mixture-of-Experts (MoE) models, which reflects the current trend in model architecture design. For dense model experiments, we utilize a dedicated testbed consisting of 64 NVIDIA H100 GPUs hosted by our research cluster and develop a custom AWS OFI NCCL plugin to restrict bandwidth between GPUs located in different emulated buildings.
The default oversubscription ratio across the emulated buildings is 1:4. 


We use testbed experiments to systematically characterize the impact of parallelism placement and scheduling. Our study covers bandwidth oversubscription ratios of 1:1, 1:2, 1:4, 1:8, and 1:16, encompassing all scenarios described in Table~\ref{table:cross_layers}. The 1:1 ratio serves as the baseline configuration of the research cluster testbed, representing no oversubscription and providing 400 Gbps of bandwidth between each pair of GPUs located in different servers. For parallelism scheduling, we evaluate FSDP and HSDP with hierarchical aggregation, as well as the effects of different pipeline schedules using DoraPP and Interleaved ZBV. To ensure that the emulation takeaways apply to much larger scale production settings, we configure the workload to use model architecture similar to production training jobs. The detailed workload configurations are listed in Table~\ref{table:workload_configs}. 



\subsection{Simulation Settings}
\label{app:method_simulator}
\begin{table}[t]
\caption{Overview of network topologies used in simulations. "1:1" stands for no bandwidth oversubscription. The circled number in the "Configuration Space" column corresponds to the circled number in Figure~\ref{fig:scale_across_infra} and represents the scale-across network configuration we simulate with each topology. 
}
\centering
\begin{tabular}{|>{\raggedright\arraybackslash}p{1.2cm}|>{\raggedright\arraybackslash}p{2.8cm}|>{\raggedright\arraybackslash}p{2.5cm}|>{\raggedright\arraybackslash}p{2.3cm}|}
\hline
 Topology Name & Intra-building Oversubscription & Per Building Device Count & Configuration Space\\ \hline
A1 & Default & Default & \textcircled{\raisebox{-0.9pt}{1}}-\textcircled{\raisebox{-0.9pt}{4}}
\\ \hline
A2 & 2$\times$ Default & Default & \textcircled{\raisebox{-0.9pt}{1}}-\textcircled{\raisebox{-0.9pt}{3}}
\\ \hline
A3 & 4$\times$ Default & Default & \textcircled{\raisebox{-0.9pt}{1}}-\textcircled{\raisebox{-0.9pt}{3}}
\\ \hline
A4 & Default & 0.8$\times$ Default & \textcircled{\raisebox{-0.9pt}{1}},\textcircled{\raisebox{-0.9pt}{2}},\textcircled{\raisebox{-0.9pt}{4}}
\\ \hline
B1 & Default & Default & \textcircled{\raisebox{-0.9pt}{1}}-\textcircled{\raisebox{-0.9pt}{4}}
\\ \hline
B2 & 1:1 & Default & \textcircled{\raisebox{-0.9pt}{1}}-\textcircled{\raisebox{-0.9pt}{3}}
\\ \hline
B3 & 2$\times$ Default & Default & \textcircled{\raisebox{-0.9pt}{1}}-\textcircled{\raisebox{-0.9pt}{3}}
\\ \hline
B4 & Default & 0.75$\times$ Default & \textcircled{\raisebox{-0.9pt}{1}},\textcircled{\raisebox{-0.9pt}{2}},\textcircled{\raisebox{-0.9pt}{4}}
\\ \hline
\end{tabular}
\label{table:topology_configs}
\end{table}

We cover parallelism placement, communication pattern, and pipeline schedules with the testbed emulations. To fully characterize the optimization space---including the effects of link latency, packet loss rate, and network protocol design---and to enable large-scale evaluation of cross-building network impacts, we employ an in-house packet-level network simulator~\citep{arcadia}. This simulator provides a precise modeling of the end-to-end traversal of individual data packets, capturing the behavior of network interface cards (NICs) and switches. It also supports the implementation of custom load balancing and congestion control protocols. 

Simulations are configured using model architectures and workload settings that precisely mirror the 100K GPUs training job we discussed in Section~\ref{background:experience_100K}. We simulate multiple data center buildings and set the default oversubscription ratio based on that of an internal multi-building region (referred to as the "default ratio"). The default latency is 50 $\mu$s, which is the latency of 10 kilometer links and is reasonable for cross-building but intra-region connections. To extrapolate to future scenarios where millions of GPUs spanning across regions located longer than 100 kilometers apart, our simulations also consider oversubscription ratios of 2$\times$ and 4$\times$ the default, as well as link latencies of 10, 25, 100, and 1000 $\mu$s. This topology is designated as "A1" in Table~\ref{table:topology_configs}. To demonstrate the generalizability of our findings, we further simulate multiple variants of A1, along with an additional series of multi-tier Clos topologies labeled B1–B4. The most notable distinction between the A and B series is that the B series, by default, also has intra-building oversubscription.

\end{document}